\documentclass[reprint,aps,prd,nofootinbib, floatfix]{revtex4-1}
\setcitestyle{authoryear,round}
\usepackage[T1]{fontenc}
\usepackage{ae,aecompl}
\usepackage{graphicx}
\usepackage{amsmath}
\usepackage{amssymb}
\usepackage{txfonts}
\usepackage{slashed}
\usepackage{makecell}
\usepackage{capt-of}

\usepackage{caption}
\usepackage[colorlinks, citecolor=blue, linkcolor=blue]{hyperref}

\usepackage[dvipsnames]{xcolor}
\definecolor{armygreen}{rgb}{0.0, 0.6, 0.4}

\usepackage{etoolbox}

\newcommand{\ltsima}{$\; \buildrel < \over \sim \;$}
\newcommand{\lsim}{\lower.5ex\hbox{\ltsima}}
\newcommand{\gtsima}{$\; \buildrel > \over \sim \;$}
\newcommand{\gsim}{\lower.5ex\hbox{\gtsima}}

\newcommand{\Norm}{\operatorname{N}}
	
\usepackage[section]{placeins}
\usepackage[lined,algoruled,noline]{algorithm2e}
\usepackage{cleveref}

\begin{document}

\title[3D cosmic shear and 3D lensing random fields]{3D cosmic shear: numerical challenges, 3D lensing random fields generation and Minkowski Functionals for cosmological inference}

\medskip

\author{A.~Spurio~Mancini}
\affiliation{Institut f\"ur Theoretische Physik, Universit\"at Heidelberg, Philosophenweg 12, 69120 Heidelberg, Germany}
\email{spuriomancini@thphys.uni-heidelberg.de}
\author{P.~L.~Taylor}
\affiliation{Mullard Space Science Laboratory, University College London, Holmbury St.~Mary, Dorking, Surrey RH5 6NT, UK}
\author{R.~Reischke}
\affiliation{Astronomisches Rechen-Institut, Zentrum f{\"u}r Astronomie der Universit{\"a}t Heidelberg, Philosophenweg 12, 69120 Heidelberg, Germany, \\ Institut f\"ur Kernphysik, Karlsruher Institut f\"ur Technologie, 76344 Eggenstein-Leopoldshafen, Germany}
\author{T.~Kitching}
\affiliation{ Mullard Space Science Laboratory, University College London, Holmbury St.~Mary, Dorking, Surrey RH5 6NT, UK}
\author{V.~Pettorino}
\affiliation{AIM, CEA, CNRS, Universit\'{e} Paris-Saclay, Universit\'{e} Paris Diderot, Sorbonne Paris Cit\'{e}, F-91191 Gif-sur-Yvette, France}
\author{B.~M.~Sch\"{a}fer}
\affiliation{Astronomisches Rechen-Institut, Zentrum f\"ur Astronomie der Universit\"at Heidelberg, Philosophenweg 12, 69120 Heidelberg, Germany}
\author{B.~Zieser}
\author{Ph.~M.~Merkel}
\affiliation{Institut f\"ur Theoretische Astrophysik, Zentrum f\"ur Astronomie der Universit\"at Heidelberg, Philosophenweg 12, 69120 Heidelberg, Germany}

\label{firstpage}
\captionsetup{justification=raggedright,singlelinecheck=false}
\setlength{\parindent}{10pt}

\begin{abstract}
Cosmic shear - the weak gravitational lensing effect generated by fluctuations of the gravitational tidal fields of the large-scale structure - is one of the most promising tools for current and future cosmological analyses. The spherical-Bessel decomposition of the cosmic shear field (``3D cosmic shear'') is one way to maximise the amount of redshift information in a lensing analysis and therefore provides a powerful tool to investigate in particular the growth of cosmic structure that is crucial for dark energy studies. However, the computation of simulated 3D cosmic shear covariance matrices presents numerical difficulties, due to the required integrations over highly oscillatory functions. We present and compare two numerical methods and relative implementations to perform these integrations. We then show how to generate 3D Gaussian random fields on the sky in spherical coordinates, starting from the 3D cosmic shear covariances. To validate our field-generation procedure, we calculate the Minkowski functionals associated with our random fields, compare them with the known expectation values for the Gaussian case and demonstrate parameter inference from Minkowski functionals from a cosmic shear survey. This is a first step towards producing fully 3D Minkowski functionals for a lognormal field in 3D to extract Gaussian and non-Gaussian information from the cosmic shear field, as well as towards the use of Minkowski functionals as a probe of cosmology beyond the commonly used two-point statistics.
\end{abstract}
%

\date{\today}

\maketitle

\section{Introduction}

The weak gravitational lensing effect is the distortion of light bundles due to the differential deflection of light rays coming from distant sources \citep[see e.g.][for reviews on the topic]{Bartelmann2001, Hoekstra2008, Kilbinger2015} caused by the gravitational tidal fields generated by the mass distribution between the source and the observer. 
This effect induces a small change in the ellipticity of a background galaxy's image, which is referred to as `cosmic shear' when it is caused by the gravitational fields of the large-scale structure of the Universe. Cosmic shear measurements are of a statistical nature, as the lensing effect is not associated with a particular intervening lens, but rather corresponds to small distortions (of the order of 1$\%$) by all potential fluctuations along the line of sight; detecting the extremely faint cosmic shear signal requires averaging over many background galaxies. The statistical properties of the shear field reflect those of the underlying density field by virtue of the gravitational field equations. The cosmic shear field has zero mean; at the level of one-point statistics, cosmological information can be extracted from e.g. peak counts \citep{Lin2015, Peel2017, Fluri2018, Vallis2018}, while for two-point statistics one looks in configuration space at the angular correlation function of the shear field, or its equivalent in Fourier space, the cosmic shear angular power spectrum.~At higher order, cosmic shear can also break degeneracies between the dark sector and neutrinos \citep{Peel2018}.

Since the first detections in early 2000s \citep[e.g.][]{Bacon2000, VanWaerbeke2000, Brown2003}, cosmic shear has flourished into a well-established theoretical framework and important cosmological constraints have already been derived from cosmic shear analyses of different surveys over the last decade, such as the Canada-France-Hawaii Telescope Lensing Survey \citep[CFHTLenS;][]{Heymans2013, Kitching2014, Joudaki2017} and the Kilo Degree Survey \citep[KiDS;][]{Kohlinger2017, Hildebrandt2017, VanUitert2018}. Even stronger constraints are expected in the next decade from planned Stage IV surveys such as \textit{Euclid}\footnote{\url{https://www.euclid-ec.org/}}\citep{EuclidRedBook2011} and the Large Synoptic Survey Telescope\footnote{\url{https://www.lsst.org/}}\citep[LSST;][]{LSST2009}, which will use the weak gravitational lensing effect as one of the main cosmological probes to achieve their science goals, most importantly the investigation of the accelerated expansion of the Universe through its influence on the background expansion and cosmic structure growth. Studies to ascertain the nature of the dark energy component of the Universe, to which the acceleration can be ascribed, crucially depend on the sensitivity of the cosmological probes to the growth of cosmic structures. Cosmic shear appears therefore very promising for this purpose, because it is particularly sensitive to the growth.  

Extracting information along the radial direction becomes then a necessary requirement for any cosmic shear analysis that aims at increasing the sensitivity to the growth of cosmic structure. For this reason, a standard approach \citep[``tomography'', first introduced in][]{Hu1999} for analysing a cosmic shear survey consists in calculating correlations of the lensing signal between different redshift bins, to which the observed galaxies are assigned according to their estimated (photometric) redshifts, in an attempt to reduce loss of information due to radial averaging.

Alternatively to this tomographic approach, a spherical-Bessel decomposition of the cosmic shear field \citep[``3D cosmic shear'', first introduced in][]{Heavens2003} allows for the inclusion in the analysis of the redshift information on each galaxy within the survey. This fully 3D approach has been recently studied in \cite{SpurioMancini2018} and \cite{Pratten2016} in the context of modified gravity theories. In \cite{SpurioMancini2018} and \cite{Taylor2018a} the cosmological sensivity of the 3D and tomographic approach have been compared. The code used to produce the results in \cite{Taylor2018a} has been released in \cite{Taylor2018b}\footnote{The code is available at \url{https://github.com/astro-informatics/GLaSS}. For questions on the code, please contact P. Taylor at \href{mailto:peterllewelyntaylor@gmail.com}{peterllewelyntaylor@gmail.com} }, where a study of the flat universe approximation has been performed in view of future Stage IV surveys. \cite{Taylor2018a} also discussed alternative weights to the spherical-Bessel ones, which may be easier to compute while still preserving most of the radial information.

The 3D spherical-Bessel formalism presents challenging integrals to evaluate numerically: we highlight them while reviewing the 3D cosmic shear formalism in Sec.~\ref{sec:3Dcosmicshear}. Subsequently we describe two different numerical techniques used to evaluate those integrals, namely the numerical recipes underlying the results presented in \cite{SpurioMancini2018} and \cite{Taylor2018a, Taylor2018b}: we present them in Sec.~\ref{sec:Levin} and \ref{sec:Glass}, respectively. In Sec.~\ref{sec:code_comparison} we present a comparison of 3D cosmic shear covariance matrices 
obtained with the two numerical methods.~Having two completely independent numerical techniques to tackle the 3D cosmic shear integrations, producing results in excellent agreement between them, is useful for a number of future applications that employ the simulated 3D covariance matrices. These include e.g. a cross-correlation analysis of 3D cosmic shear and galaxy clustering \citep[see][for a spherical-Bessel analysis of a spectroscopic galaxy clustering survey]{Lanusse2015}, or the development of a Bayesian Hierarchical Model for 3D cosmic shear power spectra estimation \citep[see][for a Bayesian Hierarchical Model for tomography]{Alsing2016, Alsing2017}. 
We show in Sec.~\ref{sec:random_fields} how to make use of the 3D covariance matrices to generate Gaussian random fields on the sky, showing in particular how to overcome the difficulties arising from the non-diagonality of the covariance matrices in the radial coordinate, which originates from the inhomogeneity of the lensing field along the line of sight. We test the validity of our field-generation procedure in Sec.~\ref{sec:minkowski}, where we briefly describe and then calculate the Minkowski Functionals of our generated Gaussian random fields, and compare them with their expectation values, known analytically in the Gaussian case. We also demonstrate in a first simplified case how a likelihood analysis for cosmological inference can be carried out using the estimated Minkowski Functionals. Finally, we draw our conclusions in Sec.~\ref{sec:conclusions}.

\section{3D cosmic shear}\label{sec:3Dcosmicshear}
We begin by reviewing the formalism for a fully 3D expansion of the shear field based on its spherical-Bessel decomposition, as first introduced in lensing studies by \citet{Heavens2003}. Here we follow the notation and conventions of \citet{SpurioMancini2018} and \citet{Zieser2016} \citep[see also][]{Heavens2006, Kitching2007, Ayaita2012, Grassi2014, Kitching2015, Taylor2018a, Taylor2018b}.

Information on the gravitational lensing effect is encoded in the lensing potential, a weighted projection of the gravitational potential along the line of sight. In a standard General Relativity context \citep[see][for an extended formalism valid also for modified gravity theories]{SpurioMancini2018, Pratten2016}, the lensing potential $\phi$ is related to the gravitational potential $\Phi$ by
\begin{align}\label{eq:lensing_potential}
\phi (\chi, \hat{\mathbf{n}}) = \frac{2}{c^2}\int_0^{\chi} \mathrm{d}\chi' \frac{\chi-\chi'}{\chi \chi'} \Phi(\chi, \hat{\mathbf{n}})\;, 
\end{align} 
where $\chi$ is a comoving distance, and the normalized vector $\hat{\mathbf{n}}$ selects a direction on the sky. Here and throughout the paper spatial flatness will be assumed \citep[for expressions for a non-flat Universe, see][]{Taylor2018b}, and the integration in Eq.\ref{eq:lensing_potential} is carried out in Born approximation, i.e. along the unperturbed light path. The shear tensor $\gamma (\chi, \hat{\mathbf{n}})$ is defined as the second $\slashed{\partial}$-derivative \citep{Newman1962, Goldberg1967} of the lensing potential \citep{Heavens2003, Castro2005}
\begin{align}\label{eq:edth-derivative}
\gamma (\chi, \hat{\mathbf{n}}) = \frac{1}{2} \eth \eth \phi (\chi, \hat{\mathbf{n}})\;.
\end{align}
The $\eth$-derivative acts as a covariant differentiation operator on the celestial sphere and relates quantities of different spin, raising the spin $s$ of a field, a number which characterises its transformation properties under rotations. Acting twice on $\phi$, the $\eth$ operator relates the scalar (spin-0) lensing potential $\phi$ to the spin-2 shear field $\gamma$.
The shear $\gamma$ can be expanded with a choice of basis functions given by a combination of spherical Bessel functions $j_\ell(z)$ \citep{Abramovitz1988} and spin 2-weighted spherical harmonics $_2 Y_{\ell m}(\hat{\mathbf{n}})$ 
\begin{align}\label{eq:sphericalFourier-Bessel}
\gamma (\chi, \hat{\mathbf{n}}) = \sqrt{\frac{2}{\pi}} \sum_{\ell m} \int k^2 \, \mathrm{d} k \, \gamma_{\ell m}(k) \, _2 Y_{\ell m} (\hat{\mathbf{n}}) \, j_\ell(k \chi)\;,
\end{align} 
where the coefficients $\gamma_{\ell m}(k)$ are given by
\begin{align}\label{eq:coefficients_gamma}
\gamma_{\ell m}(k) = \sqrt{\frac{2}{\pi}} \int \chi^2 \mathrm{d}\chi \int \mathrm{d}\Omega \, \gamma(\chi, \hat{\mathbf{n}}) \, j_\ell(k \chi) \, _2 Y_{\ell m}^*(\hat{\mathbf{n}})\;.
\end{align}
Inserting Eqs.\ref{eq:lensing_potential} and \ref{eq:edth-derivative} in \ref{eq:coefficients_gamma}, and applying a spherical-Bessel expansion to the gravitational potential $\Phi$, we can rewrite $\gamma$ as
\begin{align}
&\gamma(\chi, \hat{\mathbf{n}}) = \sqrt{\frac{2}{\pi}} \frac{1}{c^2} \int_0^{\chi} \mathrm{d}\chi' \, \frac{\chi-\chi'}{\chi \chi'} \\
&\times \int k^2 \mathrm{d}k  \sum_{\ell m} \sqrt{\frac{(\ell+2)!}{(\ell-2)!}} \Phi_{\ell m}(k, \chi') j_\ell(k \chi') \, _2Y_{\ell m}(\hat{\mathbf{n}})\;.\nonumber
\end{align}
 
Poisson's equation can be used to link the coefficients in the spherical-Bessel decomposition of the lensing potential to those of the density contrast field $\delta_{\ell m}(k, \chi)$,
\begin{align}
\frac{\Phi_{\ell m}(k,\chi)}{c^2} = - \frac{3}{2} \frac{\Omega_m}{(k \chi_H)^2}\frac{\delta_{\ell m}(k,\chi)}{a(\chi)}\;,
\end{align}
with the Hubble radius $\chi_H \equiv c/H_0$. If fluctuations in $\Phi$ are weak, the density field is statistically homogeneous and isotropic, characterised by a power spectrum $P_\delta(k, z, z')$ which is diagonal in harmonic space
\begin{align}
\left\langle \delta_{lm}(k, z) \delta_{\ell'm'}^{*}(k', z') \right\rangle = \frac{P_{\delta}(k, z, z')}{k^2}\delta^D(k-k') \delta_{\ell \ell'}^K \delta_{mm'}^K\;.
\end{align}

Using this, we can relate the covariance of shear modes to the matter power spectrum by
\begin{align}\label{eq:3Dcovariance}
\left\langle\bar{\gamma}_{lm}(k)\bar{\gamma}_{\ell'm'}^*(k')\right\rangle &= \frac{9 \Omega_m^2}{16 \pi^4 \chi_H^4}\frac{(\ell+2)!}{(\ell-2)!}\\
&\times \int \frac{\mathrm{d}\tilde{k}}{\tilde{k}^2} \, G_\ell(k, \tilde{k}) \, G_\ell(k', \tilde{k}) \, \delta_{\ell \ell'}^K \, \delta_{m m'}^K\;,\nonumber
\end{align} 
where
\begin{align}
G_\ell(k,k') &= \int \mathrm{d}z \, n_z(z) \, F_\ell(z,k) \, U_\ell(z,k')\;,\label{eq:G}\\ 
F_\ell(z,k) &= \int \mathrm{d}z_p \, p(z_p|z) \, j_\ell[k \chi^0(z_p)]\;,\label{eq:F}\\ 
U_\ell(z,k) &= \int_0^{\chi(z)} \frac{\mathrm{d} \chi'}{a(\chi')} \frac{\chi-\chi'}{\chi \chi'} j_\ell(k \chi') \, \sqrt{P_\delta \left( k, z \left( \chi \right) \right)}\label{eq:U}\;. \\ \nonumber
\end{align}
$\bar{\gamma}$ are estimates of the shear modes that, in addition to the pure lensing effect, keep into account the redshift distribution of the lensed galaxies $n_z(z)$ and the conditional probability $p(z_p|z)$ of estimating the redshift $z_p$ given the true redshift $z$.~The contribution to the total signal coming from sources situated at different distances is governed by the source density $n_z(z)$; through this term, the survey depth affects the strength of the overall signal.~Angular variations are assumed to be negligible by considering a uniform source density: the number of sources per steradian and redshift interval is approximated by the mean $n_z (z)/(4\pi)$ across the sky. The influence of incomplete sky coverage is ignored in this formalism: for applications to a Fisher matrix analysis, for example, the effect of partial sky coverage can be well compensated by a multiplying factor $f_{\mathrm{sky}}$ denoting the fraction of sky spanned by the survey, prepended to the expression for the Fisher matrix~\citep{Heavens2006, SpurioMancini2018}.~A finite field of view can be incorporated in the analysis by considering a suitable window function $W(\hat{\mathbf{n}})$ that represents the angular distribution of the sources (e.g. a top hat filter corresponding to a rectangular field of view). For details on the extended formalism to include in the analysis this inhomogeneous sampling we refer the reader to e.g. \citet{Heavens2003}, while e.g. \citet{Leistedt2015} consider alternative methods, such as wavelets, to deal with survey geometry in 3D cosmic shear.

Statistical isotropy guarantees that the covariance in Eq.~\ref{eq:3Dcovariance} does not depend on the multipole order $m$, while the assumed full sky coverage also prevents mixing of different $\ell$ modes.~If the finite field of view is taken into account, statistical isotropy is broken (e.g. by the absence of data across parts of the sky) leading to a coupling of different $\ell$-modes; furthermore, if the field of view is not square, even for a fixed $\ell$ there will be different results for different $m$-modes. The lensing weight function, the redshift errors and the redshift-dependence of the source distribution, instead, always introduce correlations between the amplitudes of the signal on different scales; the covariance matrix then acquires off-diagonal terms, the calculation of which is numerically involved \citep[see][for how to take these into account using a pseudo-$C_\ell$ approach in 3D]{Kitching2014}. 

The basis of spherical Bessel functions leads to integrals with rapidly oscillatory kernels, which in the inference process have to be solved for a large number of parameter combinations. The $\sqrt{P_\delta \left( k, z\left( \chi \right) \right)}$ term comes from an approximation, introduced and qualitatively justified in \citet{Castro2005}, to calculate unequal-time correlators appearing in the comoving distance integrations by means of a geometric mean $P \left( k, z, z' \right) \simeq \sqrt{P \left( k, z \right) P \left( k, z' \right)}$  \citep[see also][]{Kitching2017}. This expression simplifies considerably in the linear regime of structure formation, retrieving the one presented in the seminal paper of \citet{Heavens2003} where a product of the linear growth factors at different redshifts is present, acting on the matter power spectrum evaluated at the present time. 

The noise term for the covariance matrix of the shear modes is given by the intrinsic ellipticity dispersion of source galaxies, as a result of the fact that the observed ellipticity $\epsilon$ is assumed to be the sum of the shear $\gamma$ and the intrinsic ellipticity $\epsilon_S$. The intrinsic ellipticity dispersion is given by $\left\langle \epsilon_S^2 \right\rangle = \sigma_\epsilon^2$. As given in \citet{Heavens2006} and \citet{Kitching2007}, in the spherical-Bessel formalism (see Appendix~\ref{app:noise} for a complete derivation), this leads to
\begin{align}\label{eq:noise}
\left\langle \gamma_{\ell m}\left( k \right) \gamma_{\ell' m'}\!\left( k' \right) \right\rangle_{\mathrm{SN}}\!=\! \frac{\sigma_\epsilon^2}{2 \pi^2}\!\! \int\!\!\mathrm{d}z \, n_z(z) j_\ell \left[ k \chi_0(z) \right] j_{\ell'} \left[ k' \chi_0(z) \right] \delta_{\ell \ell'}^K \delta_{m m'}^K\; ,
\end{align}
where the subscript SN stands for ``shot noise'' and we set $\sigma_\epsilon = 0.3$. This expression for the noise holds only in absence of intrinsic alignments, i.e. assuming that the intrinsic ellipticities of galaxies are uncorrelated \cite[see][for a study of intrinsic alignments in 3D cosmic shear]{Merkel2013, Kitching2015}.

\section{Numerical Implementation}
In this section we will briefly describe the two methods used to calculate the correlations from Eqs. (\ref{eq:3Dcovariance}) and (\ref{eq:noise}). While one code implements in C\texttt{++} the Levin collocation method \citep{Levin1996, Levin1997} that makes use of the periodic oscillations of the Bessel functions and has been used to produce the results of \citet{Zieser2016} and \citet{SpurioMancini2018}, the other implements the integrations by matrix multiplications and appropriate use of the Limber approximation \citep{Kaiser1992, Kaiser1998, Loverde2008} at high $\ell$. The second code is a \texttt{Python} module, implemented in the code {\tt GLaSS} and used in \citet{Taylor2018a, Taylor2018b}.

\subsection{Levin integration}\label{sec:Levin}
The method presented in \citep{Levin1996, Levin1997} can be used for efficient evaluation of rapidly oscillatory integrals, once certain conditions are satisfied. The main idea behind the method is to transform the quadrature problem into the solution of a system of linear ordinary differential equations. These are then tackled by collocation, i.e. choosing candidate solutions (polynomials) and a number of points in the domain (called collocation points), and selecting that solution which satisfies the given equations at the collocation points.

As seen in Sec.~\ref{sec:3Dcosmicshear} the 3D cosmic shear signal and noise (cf.~Eqs.~\ref{eq:3Dcovariance} and \ref{eq:noise}) present several integrals of the form
\begin{align}\label{eq:i1}
I_1 [h] = \int_{z_1}^{z_2} \mathrm{d} z \,  h \left( z \right) j_\ell \left( k \chi \left( z \right) \right)
\end{align}
or  
\begin{align}\label{eq:i2}
I_2 [h] = \int_{z_1}^{z_2} \mathrm{d} z \, h \left( z \right) j_\ell \left( k_1 \chi \left( z \right) \right) j_\ell \left( k_2 \chi \left( z \right) \right).
\end{align}
The comoving distance between two events at redshift $z_1$ and $z_2$ is given by
\begin{align}
\chi (z_1, z_2) = \int_{a(z_2)}^{a(z_1)} \frac{c \mathrm{d}a}{a^2 H(a)} = \chi_H \int_{z_1}^{z_2} \frac{\mathrm{d}z}{E(z)}\;,
\end{align}
where $a$ is the scale factor and $\dot{a}/a = H(a) = H_0 E(a)$ is the Hubble function.
Rather than redshift integrals, Eqs.~\ref{eq:i1} and \ref{eq:i2} can be rewritten, using $\mathrm{d} z = \mathrm{d} \chi \, E[z(\chi)]$, with the comoving distance as the integration variable:
\begin{align}
I_1 [h] &= \frac{1}{\chi_H} \int_{\chi(z_1)}^{\chi(z_2)} \mathrm{d} \chi \, E[z(\chi)] \, h[z(\chi)] \, j_\ell (k \chi) \label{1bessel}\;, \\
I_2 [h] &= \frac{1}{\chi_H} \int_{\chi(z_1)}^{\chi(z_2)} \mathrm{d} \chi \, E[z(\chi)] \, h[z(\chi)] \, j_\ell (k_1 \chi) j_\ell (k_2 \chi)\;. \label{2bessel}
\end{align}
Due to the highly oscillatory nature of the spherical Bessel functions, especially at high $\ell$ or $k$, the numerical solution of these integrals by standard quadrature routines is extremely inaccurate when a large number of zero-crossings occurs in the interval $[\chi(z_1), \chi(z_2)]$, unless an enormous number of points is used to sample the integrand: however, the procedure becomes then exceedingly time-consuming, especially if many combinations of $\ell$ and $k$ need to be considered.

Here we describe an alternative method, presented by \cite{Levin1996, Levin1997}, which we use to evaluate our integrals. It is applicable to integrals of the form
\begin{align}\label{scalarprod}
I[F] = \int_a^b \mathrm{d} x \, F^T (x) w(x)  = \int_a^b \mathrm{d} x \left\langle F, w \right\rangle (x)\;	,
\end{align}
where $F(x) = [F_1(x), \ldots, F_d(x)]^T$ and $w(x) = [w_1(x), \ldots, w_d(x)]^T$ are vectors of functions, for which the second equality of Eq.~\ref{scalarprod} defines a scalar product $\left\langle , \right\rangle$ and the functions $w_i(x), i = 1, 2, . . . , d$, but not $F_i(x)$, are rapidly oscillatory across the integration domain. A matrix of functions $A(x)$ is defined, such that the derivatives of $w(x)$, denoted by $w'(x)$, fulfil
\begin{align}\label{derivata}
w'(x) = A(x) w(x)\;.
\end{align}
The components $A_{iq}(x)$ should not be highly oscillatory. We show below an example of such a matrix for the particular cases given in Eqs.~\ref{1bessel} and \ref{2bessel}. In the Levin formalism a vector $p(x)$ is constructed to approximate the integrand in Eq.~\ref{scalarprod} by
\begin{align}
\left\langle p, w \right\rangle' = \left\langle p' + A^T p , w \right\rangle \approx \left\langle F, w \right\rangle\;.
\end{align}
The first equality follows from applying the Leibniz rule for derivatives and Eq.\ref{derivata}, with $\left\langle p, A w \right\rangle = \left\langle A^T p, w \right\rangle$. If such a vector is found, then the integral in Eq.\ref{scalarprod} can be approximated by
\begin{align}
I[F] \approx \int_a^b \mathrm{d}x \left\langle p, w \right\rangle ' (x) = \left\langle p, w \right\rangle (b) - \left\langle p, w \right\rangle (a)\;.
\end{align}
This can be achieved by demanding that both terms should be equal, $\left\langle p, w \right\rangle' = \left\langle F, w \right\rangle$, at $n$ collocation points $x_j , j = 1, 2, . . . , n$. The requirement 
\begin{align}
\left\langle p' + A^T p - F, w \right\rangle (x_j) = 0, \quad j = 1, ..., n
\end{align}
generally means that the vector $\left\langle p' + A^T p - F \right\rangle$ must be orthogonal to $w$ at the points $x_j$, for example by demanding that it should be the null vector:
\begin{align}\label{zeros}
p'(x_j) + A^T (x_j) p(x_j) = F(x_j).
\end{align}
Finding a vector $p$ which has this property can be achieved by choosing a set of $n$ linearly independent and differentiable basis functions $u_m(x)$ and writing each component $p_i (x)$ as a linear combination:
\begin{align}
p_i(x) = c_i^{(m)} u_m(x), \quad i= 1,...,d; \quad m = 1,...,n\;.
\end{align}
Equation \ref{zeros} then leads to the following linear system of equations for the $d \times n$ coefficients $c_i^{(m)}$:
\begin{align}\label{740}
c_i^{(m)}u_m'(x_j) + A_{q i} c_q^{(m)} u_m(x_j) = F_i(x_j), \quad i,q = 1, ...,d; j,m = 1, ...,n\;.
\end{align}
\citet{Levin1996} showed how to concretely apply this algorithm to several cases of integrals with highly oscillatory kernels. The performance varies depending on the integrand, but accuracies below $10^{-6}$ can often be achieved with less than 10 collocation points.
As suggested by \citet{Levin1996}, in our implementation we use equidistant collocation points
\begin{align}
x_j = a + (j-1) \frac{b-a}{n-1}, \quad j = 1, ...,n 
\end{align}
and choose the $n$ lowest-order polynomials as basis functions:
\begin{align}
u_m(x) = \left( \frac{x- \frac{a+b}{2}}{b-a} \right)^{m-1}, \quad m=1, ...,n\;.
\end{align}
We note that the polynomials with $m > 1$ and the derivatives with $m > 2$ share the root $x = (a+b)/2$: to prevent the linear system of equations from becoming singular, that root should not be used as a collocation point. The factor $1/(b - a)$ is included for numerical reasons: if $b \gg 1$ or $b \ll 1$, the values of polynomials of different order may differ by several orders of magnitude; the normalising factor guarantees that $|u_m (x)|\leq 1$ across the integration domain, in order to regulate the range of the coefficients of the linear system of equations in Eq.~\ref{740} and thus the condition of the corresponding matrix. Suitable vectors $w$ for the integrals in Eqs.~\ref{1bessel} and \ref{2bessel} can be identified by considering the following recurrence relations for the spherical Bessel functions \citep{Abramovitz1988}:
\begin{align}
\frac{\mathrm{d}}{\mathrm{d}x} j_\ell(x) &= j_{\ell-1}(x) - \frac{\ell+1}{x} j_\ell (x)\;, \\
\frac{\mathrm{d}}{\mathrm{d}x} j_{\ell-1}(x)&= -j_\ell (x) + \frac{\ell-1}{x} j_{\ell-1}(x)\;.
\end{align}
Rewriting these relations in the form $w' = Aw$, one finds that
\begin{align}
w(\chi) = \left( \begin{matrix}
j_\ell(k \chi) \\
j_{\ell-1}(k \chi)
\end{matrix} \right), \quad A(\chi) = \left( \begin{matrix}
- \frac{\ell+1}{\chi} & k \\
-k & \frac{\ell-1}{\chi}
\end{matrix} \right)	 
\end{align}
is a suitable choice for the integral in Eq.~\ref{1bessel}, with $F(\chi) = \{E[z(\chi)]h[z(\chi)], 0\}^T$.~It is easy to verify that neither the entries of the matrix $A$ nor the integral kernels $F$ are rapidly oscillatory. For integrals of the type in Eq.~\ref{2bessel}, four-dimensional vectors are needed:
\begin{align}
w(\chi) = \left( \begin{matrix}
j_\ell (k_1 \chi) j_\ell(k_2 \chi) \\
j_{\ell-1} (k_1 \chi) j_\ell(k_2 \chi) \\
j_\ell (k_1 \chi) j_{\ell-1}(k_2 \chi) \\
j_{\ell-1} (k_1 \chi) j_{\ell-1}(k_2 \chi)
\end{matrix} \right), \quad A(\chi) = \left( \begin{matrix}
-\frac{2(l+1)}{\chi}\!\!\!\!\!\! & k_1\!\!\!\!\!\! & k_2\!\!\!\!\!\! &\!\!\!\!\!\! 0 \\
-k_1 & -\frac{2}{\chi} & 0 & k_2 \\
-k_2 & 0 &-\frac{3}{\chi} & k_1 \\
0 & -k_2 & -k_1 & \frac{2(\ell-1)}{\chi}
\end{matrix} \right).
\end{align}
Similarly, $F(\chi) = \{E[z(\chi)]h[z(\chi)], 0, 0, 0\}^T$.

\subsection{{\it GLaSS}}\label{sec:Glass}
\begin{figure}
\centering
\includegraphics[width=0.45\textwidth]{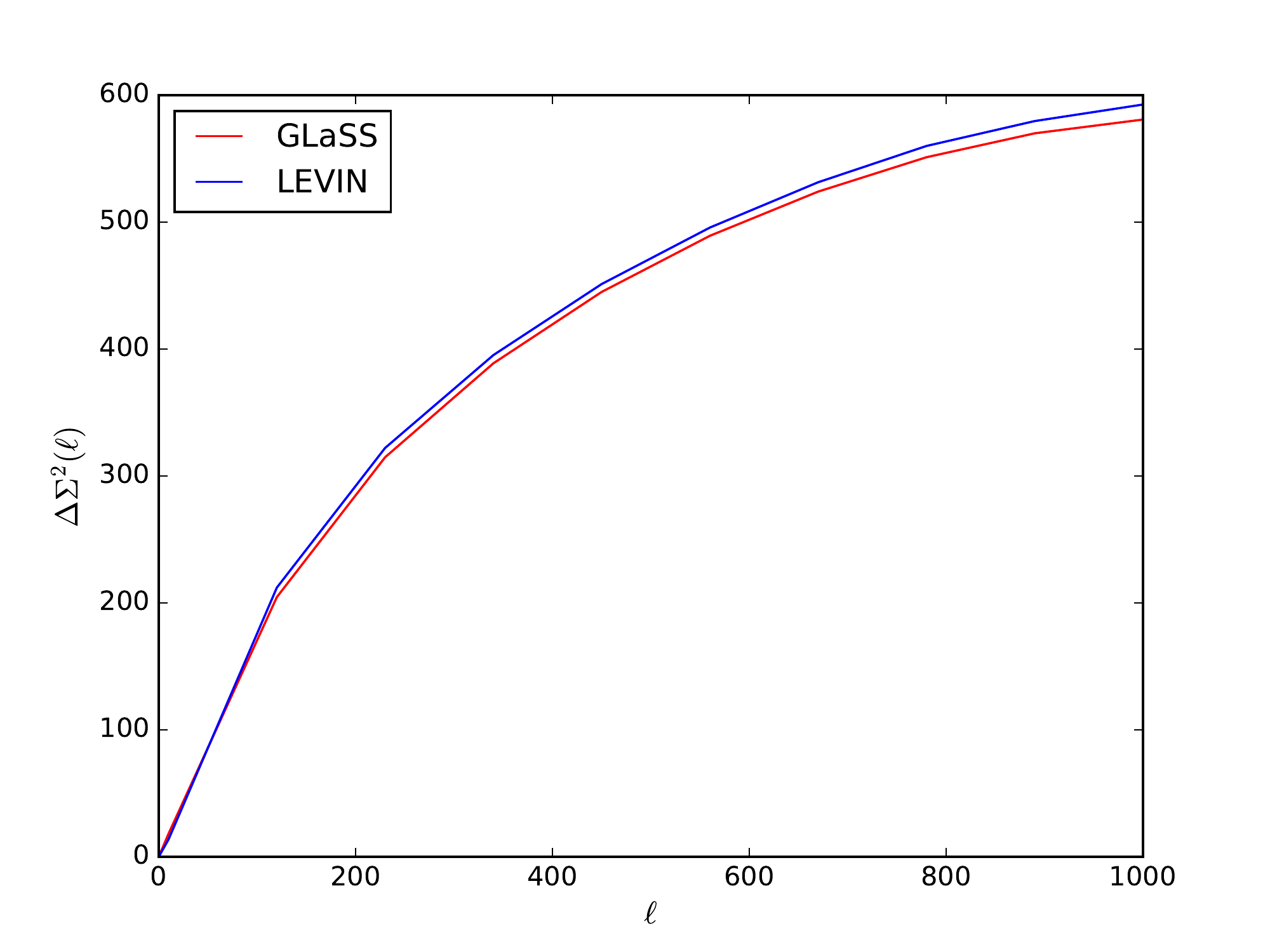}
\caption{Comparison of the differential signal-to-noise curve (eq.\ref{eq:SNR}) as a function of the angular multipole. The two curves have  been obtained from the signal and noise parts of the covariance matrices produced with {\tt GLaSS} (\textit{red}) and the Levin method (\textit{blue}).}\label{fig:snr}
\end{figure}
\begin{figure}
\centering
\includegraphics[width=0.45\textwidth]{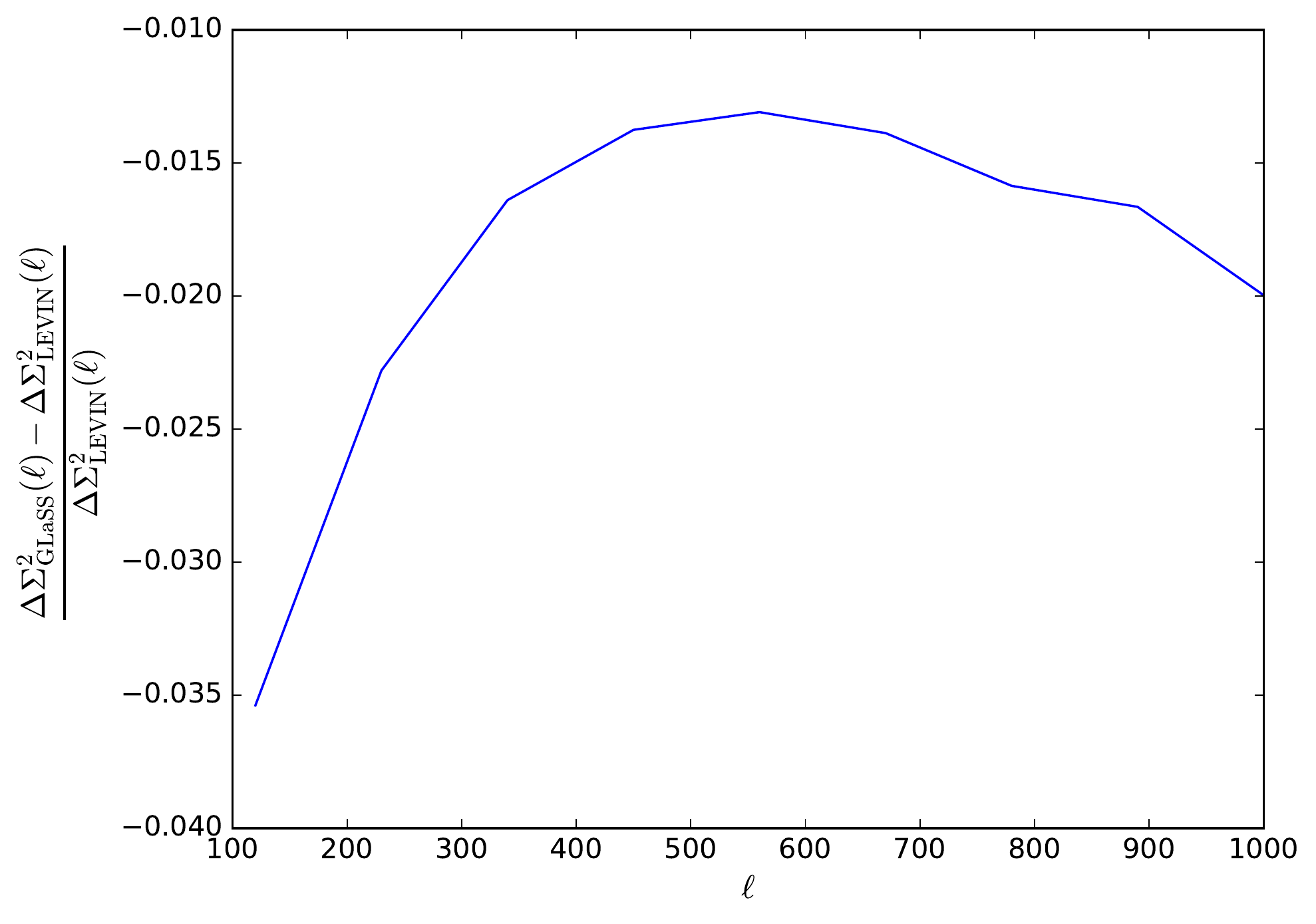}
\caption{Relative difference of the signal-to-noise curve calculated with the {\tt GLaSS} and Levin method, as a function of the multipole $\ell$.}\label{fig:snrdiff}
\end{figure}

The Generalised Lensing and Shear Spectra ({\tt GLaSS}) code is written in {\tt Python} and integrated into the modular cosmological package {\tt Cosmosis}~\citep{cosmosis}. Cosmological information can be read from an external source as in this work, or directly from the {\tt Cosmosis} pipeline. More information can be found in ~\cite{Taylor2018b}. 
\par {\tt GLaSS} is written to compute the lensing spectra for an arbitrary weight function $W_\ell[k \chi^0(z_p)]$ which takes the place of the Bessel functions in Eq.~\ref{eq:F}; see~\cite{Taylor2018a} for more details about this \textit{generalized spherical-transform}. Nevertheless, 3D cosmic shear comes as an in-built run-mode option.
\par All nested integrals in Eqs.~\ref{eq:3Dcovariance}-\ref{eq:U} are computed as matrix multiplications because this is one of the few operations that releases the Global Interpreter Lock in {\tt Python} allowing parallelisation. For example,  
\begin{equation} \label{eq:mat}
U_{\ell} \left(z, k \right) \approx   \sum_{\chi'} A \left(\chi \left(z \right), \chi' \right)B \left(\chi', k \right),
\end{equation}
$A \left( \chi, \chi' \right) \equiv \Delta \chi' \frac{F_K \left(\chi, \chi' \right)}{a \left(\chi' \right)} $, where $\Delta \chi'$ is the spacing between the sampled points in $\chi'$ and $B \left( \chi, \chi' \right) \equiv  j_{\ell} \left( k \chi' \right) \sqrt{P\left(k ; \chi' \right)}$\;. 
\par To further speed up computations all Bessel functions data is pre-computed in {\tt GLaSS}. To save memory, values of the Bessel functions $j_\ell \left(x \right)$ are stored in a 2D look up table in $\ell$ and $x$ and the $j_\ell \left(k \chi \right)$ are found as needed. This procedure was first described in~\cite{Seljak1996, kosowsky1998efficient}.  
\par While computing the lensing spectra in terms of nested matrix multiplications allows for easy parallelisation, this procedure does not efficiently sample the $z$-$k$ space as efficiently as the Levin integration. At high-$\ell$ where the Bessel functions oscillate quickly this means the lensing spectra must be evaluated at very high resolutions.  
\par To reduce the resolution at which the lensing spectra must be evaluated, {\tt GLaSS} takes the extended Limber approximation ~\citep{loverdelimber} above $\ell > 100$. This was shown to have negligible impact for stage IV surveys~\citep{Kitching2017limits}. Taking the Limber approximation, equation \ref{eq:U} can be rewritten as:
\begin{equation} \label{eq:limber}
U_{\ell} \left(\chi, k \right) = \frac{F_k \left( \chi, \nu \left( k \right) \right)}{k a \left( \nu\left( k \right)  \right)} \sqrt {\frac{\pi}{2 \left( {\ell} + 1/2 \right)}}  \sqrt{P \left( k, \nu\left( k \right)  \right)}\;,
\end{equation}
where $\nu\left( k \right)  \equiv \frac{{\ell}+ 1/2}{k}$. Meanwhile at low-$\ell$ the Bessel functions oscillate slowly and the nested integrals can be evaluated at lower resolution.

\section{Code comparison}\label{sec:code_comparison}
In the following we compare the predictions for the 3D cosmic shear covariance matrices produced with the Levin method and with the algorithm implemented in the {\tt GLaSS} code. For the code comparison we fix the fiducial cosmological model to a flat cosmology with parameters given in Tab.~\ref{tab:fiducial}. The source distribution and the redshift error probability need to be the same for the two codes. For the source distribution we follow \citet{Amendola2016} and choose 
\begin{align}
n_z(z) \propto n_0 \left( \frac{\sqrt{2}}{z_m} \right) ^3 z^2 \exp \left[ - \left( \frac{\sqrt{2}z}{z_m}\right)^{3/2}\right],
\end{align}  
where $z_m$ is the median redshift of the survey and $n_0$ is the observed redshift-integrated source density. We set $z_m = 0.9, n_0 = 30 \,  \mathrm{arcmin}^{-2}$. We take the redshift error distribution to be a Gaussian
\begin{align}\label{eq:redshift_probability}
p(z_p|z) = \frac{1}{\sqrt{2 \pi} \sigma(z)} \exp \left[ - \frac{(z_p-z)^2}{2 \sigma^2(z)} \right],
\end{align}
with a redshift-dependent dispersion 
\begin{align}
\sigma(z) = \sigma_z (1+z)\;.
\end{align}

\begin{table*}
\begin{tabular}{ccccccccc}
\hline
$\Omega_{\rm m}$ & $\Omega_{\rm b}$ & $\Omega_{\rm r}$ &  $\Omega_{k}$ & $w_0$ & $w_a$ & $\sigma_8$ & $n_{\rm s}$ & $h$ \\
\hline
0.315 & 0.0486 & $9.187 \times  10^{-5}$ &  0.0  & -1.0 & 0.0 &  0.834 &  0.962 & 0.674\\
\hline
\end{tabular}
\caption{Values of the cosmological parameters in the fiducial model assumed for the code comparison.\label{tab:fiducial}}
\end{table*}

We first compare the signal-to-noise curve. The cumulative signal-to-noise ratio, summed over the contributions at different multipoles up to a maximum multipole $\ell$, is defined as
\begin{equation}\label{eq:SNR}
\Sigma^2(\leq\ell) = f_\mathrm{sky}\sum_{\ell' = \ell_\mathrm{min}}^{\ell}\frac{2\ell' +1 }{2} \mathrm{Tr}\left[\boldsymbol{C}^{-1}_{\ell'}\boldsymbol{S}_{\ell'}\boldsymbol{C}_{\ell'}^{-1}\boldsymbol{S}_{\ell'}\right]\equiv \sum_{\ell' = \ell_\mathrm{min}}^{\ell}\Delta\Sigma^2(\ell')\; ,
\end{equation}
where $\boldsymbol{S}$ is the signal covariance (Eq.~\ref{eq:3Dcovariance}) only, while $\boldsymbol{C}$ refers to the sum of signal and shot noise, i.e. Eqs.~\ref{eq:3Dcovariance}$+$\ref{eq:noise}. The signal-to-noise curves produced by both codes are shown in Fig.\ref{fig:snr}, depicting the differential contributions to the signal-to-noise coming from the different multipoles.  
The Levin and {\tt GLaSS} predictions for the signal-to-noise curves show agreement with each other for the multipoles considered, reaching differences below 4$\%$, as evidenced by Fig.~\ref{fig:snrdiff} where the relative difference between the predictions of the two codes are shown. {\tt GLaSS} has slightly lower signal-to-noise at high $\ell$. This is because {\tt GLaSS} is not designed specifically for 3D cosmic shear and the signal-to-noise converges as the resolution of the computation grid is increased. This problem will be exacerbated as one includes higher and higher $\ell$ (e.g. $\ell=3000$ where there is still useful signal to add) because the Bessel functions oscillate more quickly. For our comparison we used 2000 $k$-modes linearly spaced between $k=0.005 \, h/\mathrm{Mpc}$ and $k=2.0 \, h/\mathrm{Mpc}$, and restricted the comparison to multipoles $\ell \leq 1000$; see \citet{Taylor2018b} for details on how much information is captured by {\tt GLaSS} at different $k$-resolutions and \citet{Taylor2018a} for a discussion of the run time at different resolutions.
 
\begin{figure}
\centering
\includegraphics[width=0.5\textwidth]{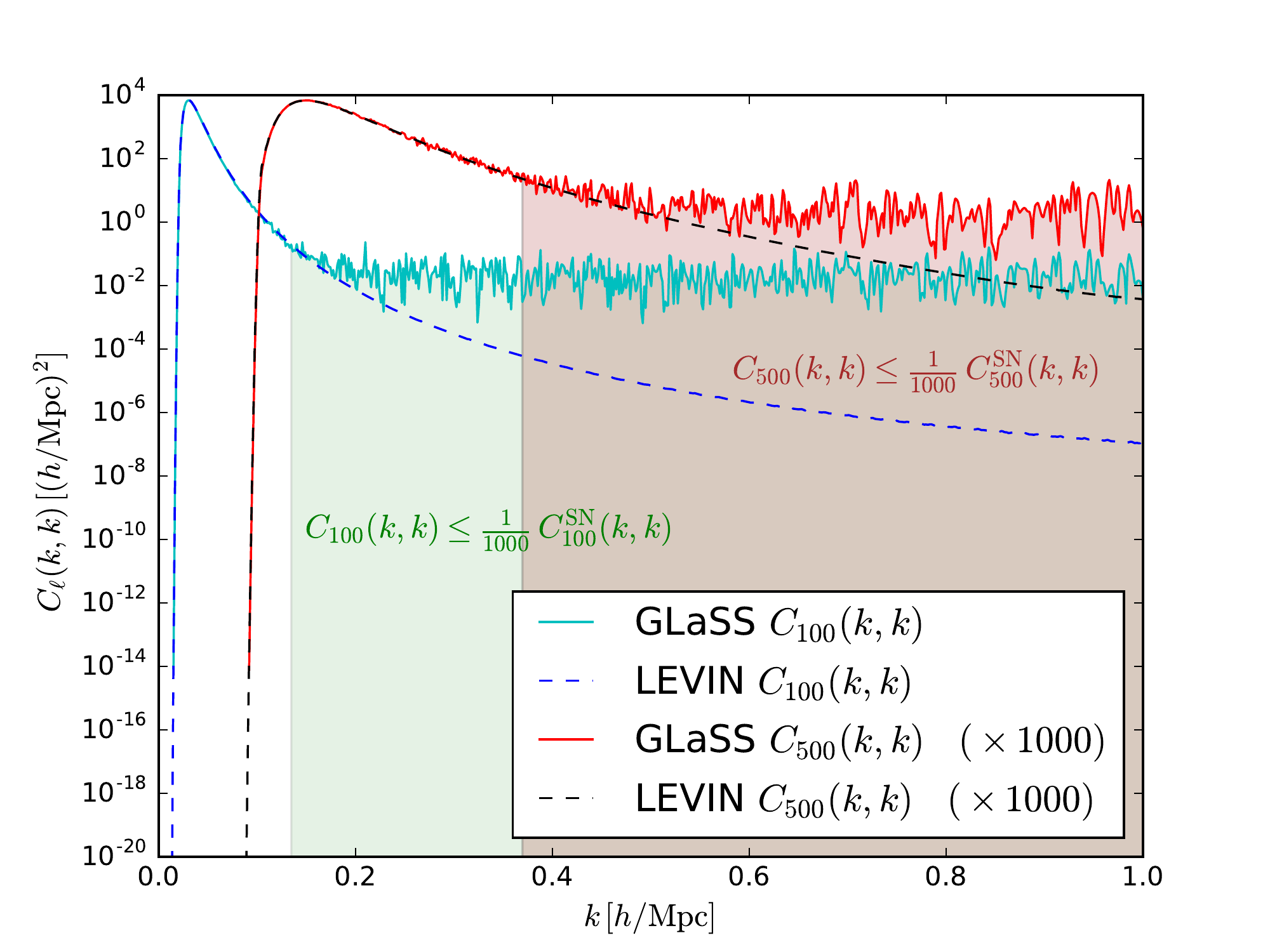}
\includegraphics[width=0.5\textwidth]{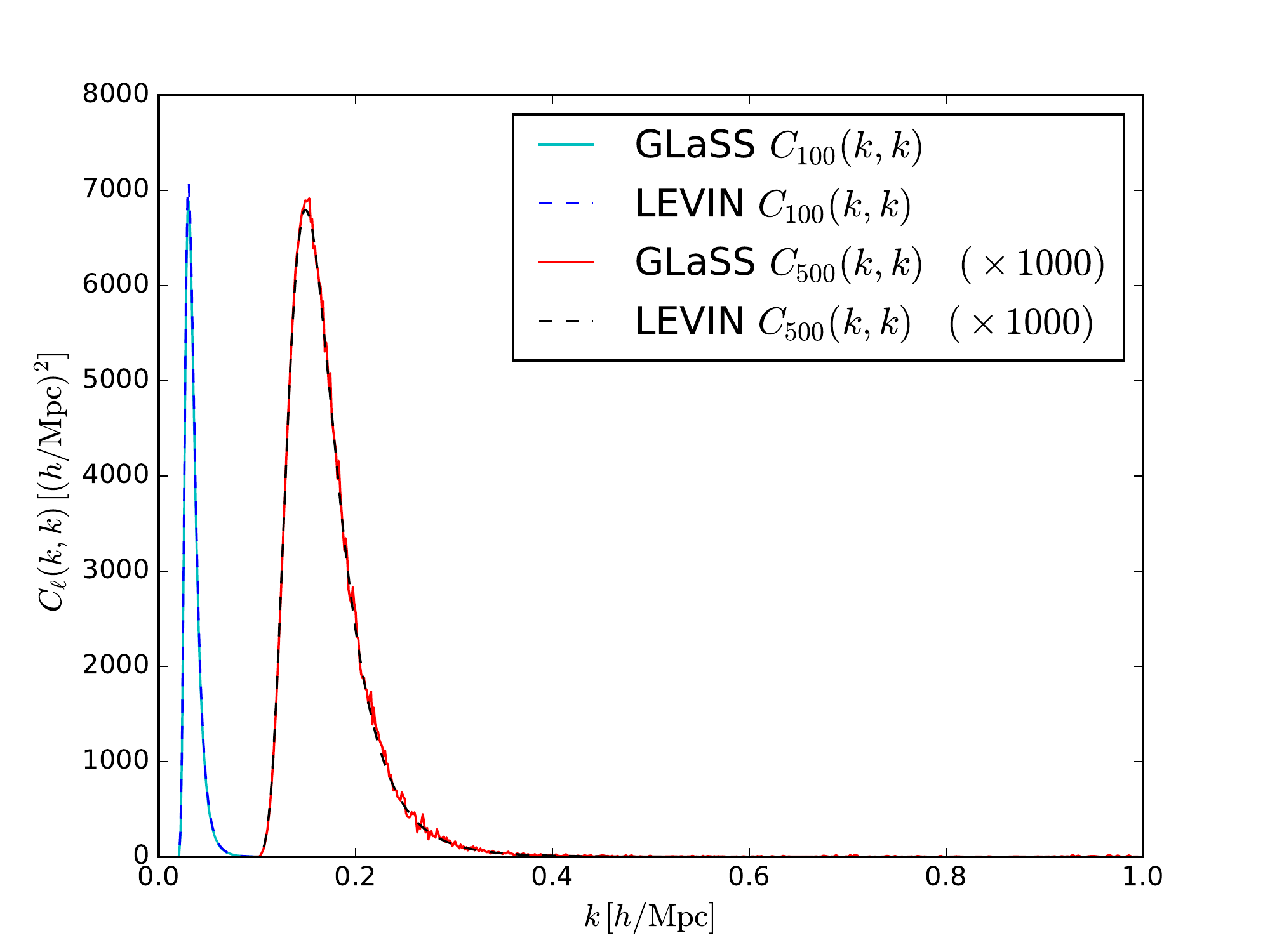}
\caption{Comparison of the diagonal elements of the signal part of the covariance matrices (eq.\ref{eq:3Dcovariance}) for two multipoles $\ell = 100$ and $\ell = 500$, produced with {\tt GLaSS} (\textit{solid} lines, \textit{cyan} and \textit{red} for $\ell = 100$ and $\ell=500$, respectively) and the Levin method (\textit{dashed} lines, \textit{blue} and \textit{black} for $\ell = 100$ and $\ell = 500$, respectively). All curves have been plotted without performing any interpolation. We show the same curves using a linear (\textit{upper} panel) and a logarithmic (\textit{bottom} panel) scale on the $y$-axis. The differences at higher $k$ ($k \gtrsim 0.2 \, h/\mathrm{Mpc}$ for $\ell =100$, $k \gtrsim 0.4 \, h/\mathrm{Mpc}$ for $\ell =500$) arise from the higher numerical noise present in the {\tt GLaSS} computations in that $k$ regime. However, these contributions are many orders of magnitude smaller than the main contributions around the peaks of the curves, and much smaller than the contributions from the noise (cf. Fig.~\ref{fig:comparison_noise}), therefore can be safely neglected. We demonstrate this point in the upper panel by indicating the shaded region for each multipole $\ell$ where the signal represents a fraction $\leq 1/1000$ of the noise: these regions correspond to the $k$-ranges where the {\tt GLaSS} and Levin predictions for the signal are in apparent disagreement (cf. also Fig.~\ref{fig:relativediff}). In both panels the curves for $\ell = 500$ have been multiplied by a factor 1000 for easier visualisation.}\label{fig:comparison_signal}
\end{figure}

\begin{figure}
\centering
\includegraphics[width=.5\textwidth]{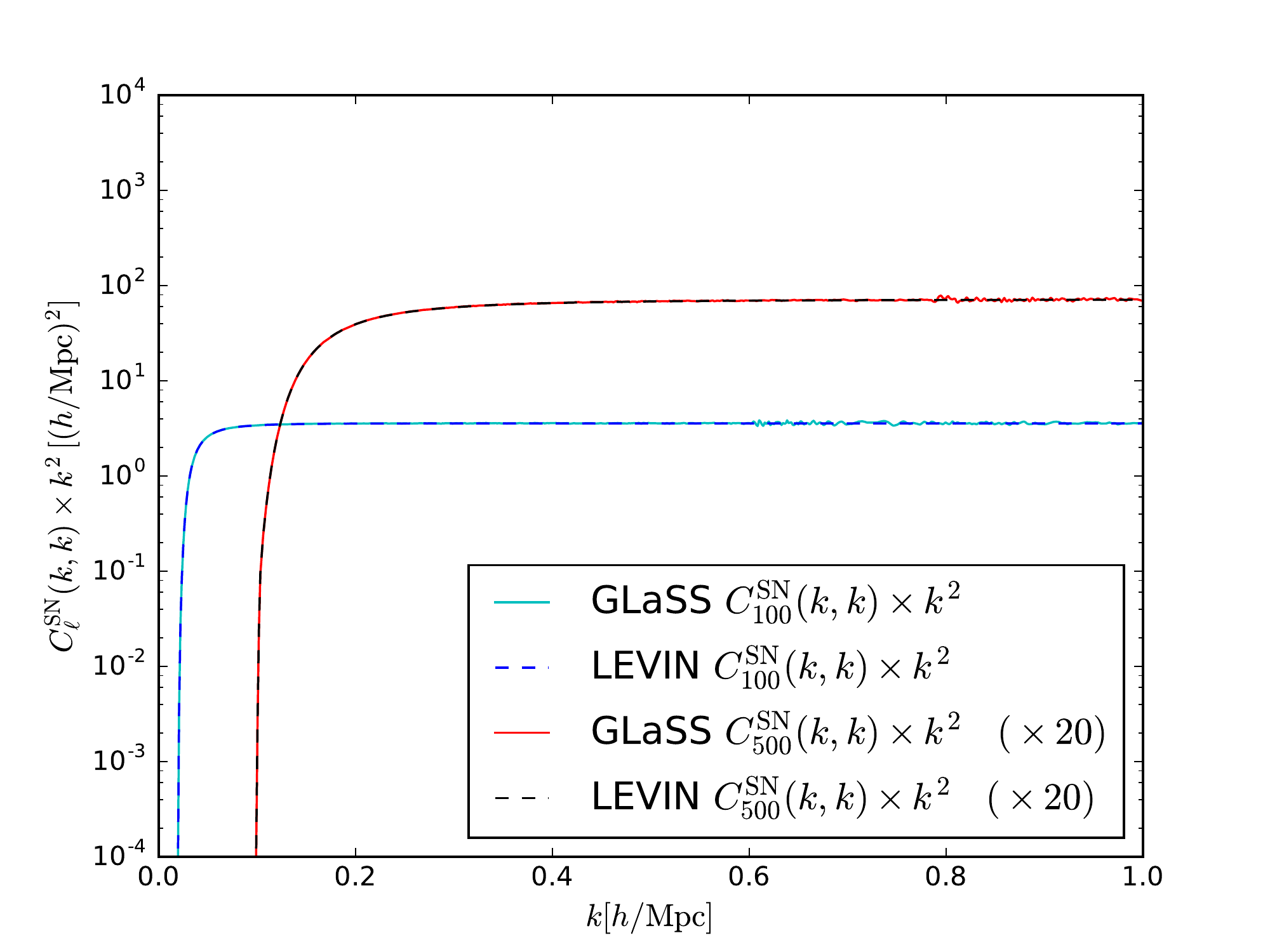}
\caption{Comparison of the diagonal elements of the noise part of the covariance matrices (eq.\ref{eq:noise}) for the same two multipoles $\ell=100$ and $\ell=500$ of Fig.~\ref{fig:comparison_signal}, produced with {\tt GLaSS} (\textit{solid}, \textit{cyan} for $\ell=100$ and \textit{red} for $\ell=500$) and the Levin method (\textit{dashed}, \textit{blue} for $\ell=100$ and \textit{black} for $\ell=500$). All curves have been plotted without performing any interpolation and multiplied by a factor $k^2$. In the case $\ell = 500$ the curves produced by both methods have also been multiplied by a factor 20 for easier visualisation.}\label{fig:comparison_noise}
\end{figure}

\begin{figure*}
\includegraphics[width=0.45\linewidth]{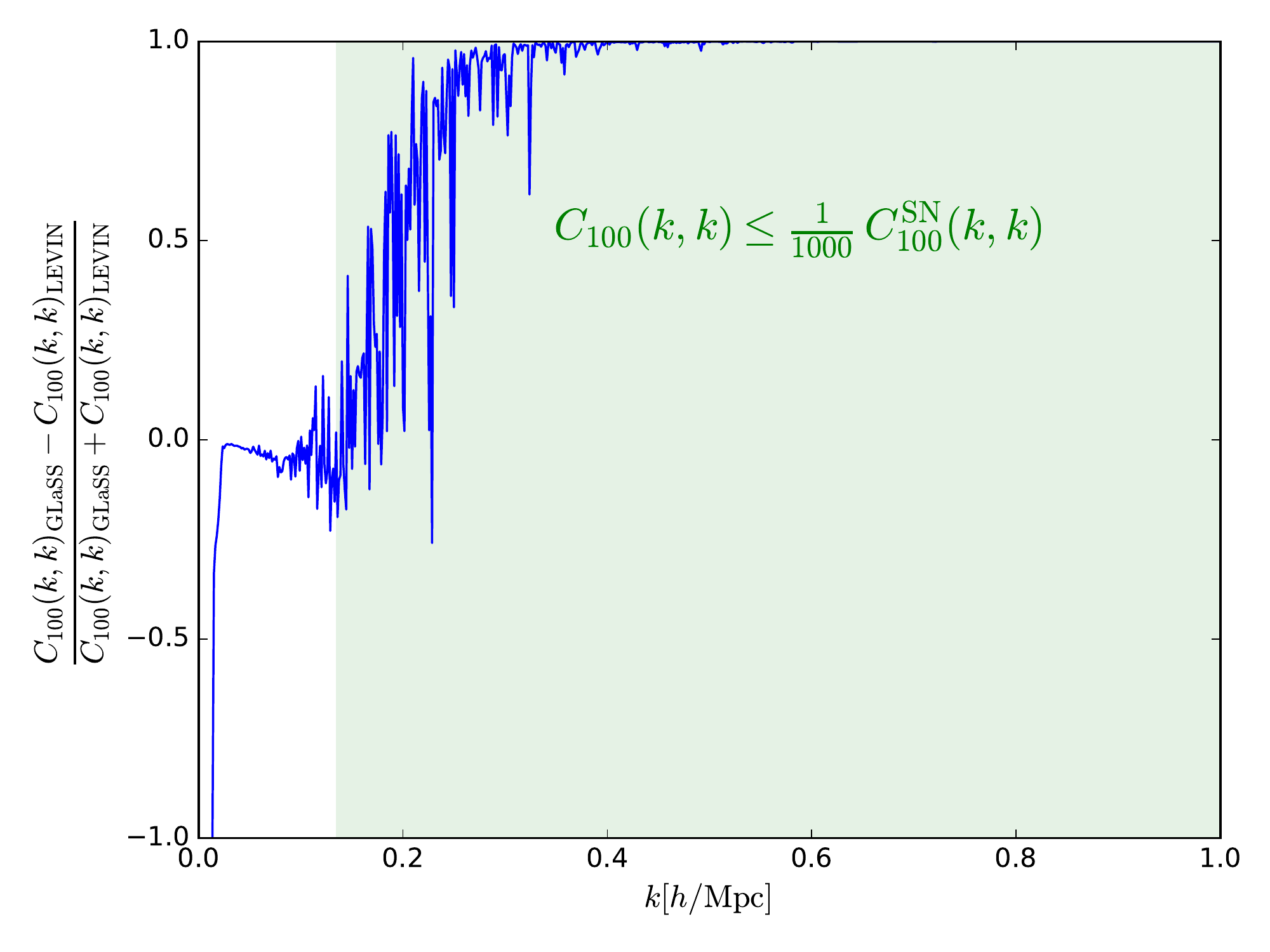}
\includegraphics[width=0.45\linewidth]{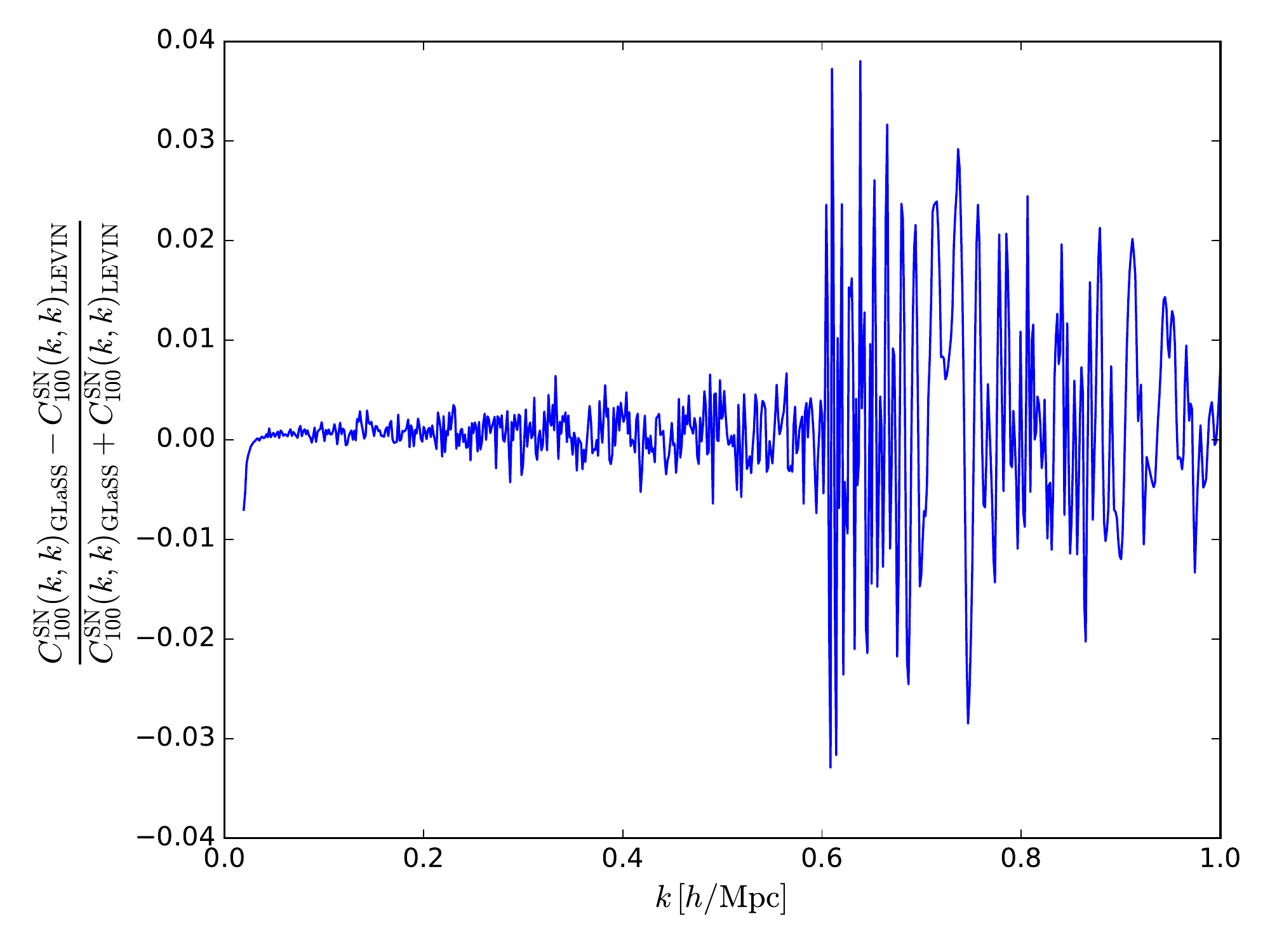}\par 
\includegraphics[width=0.45\linewidth]{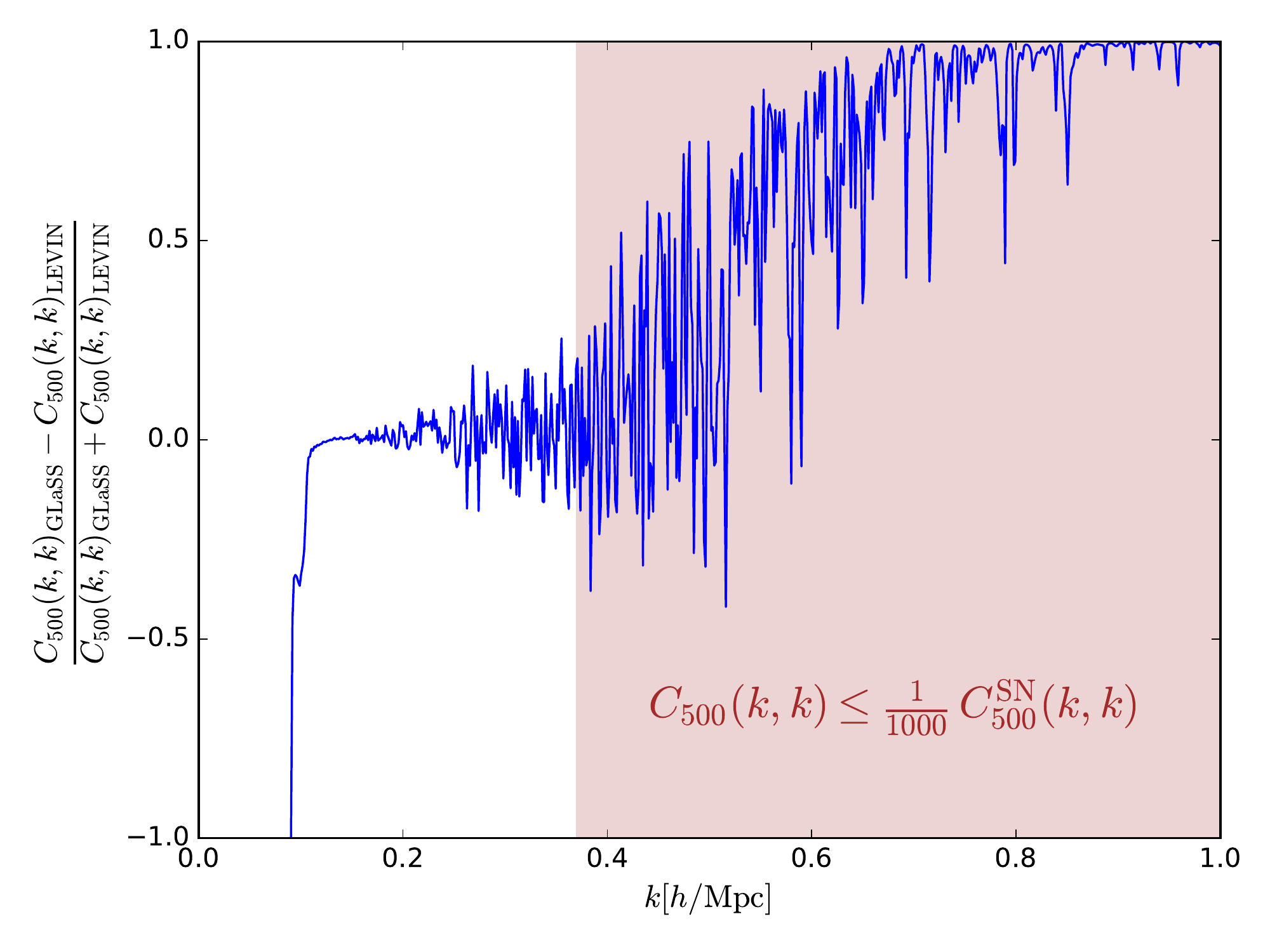}
\includegraphics[width=0.45\linewidth]{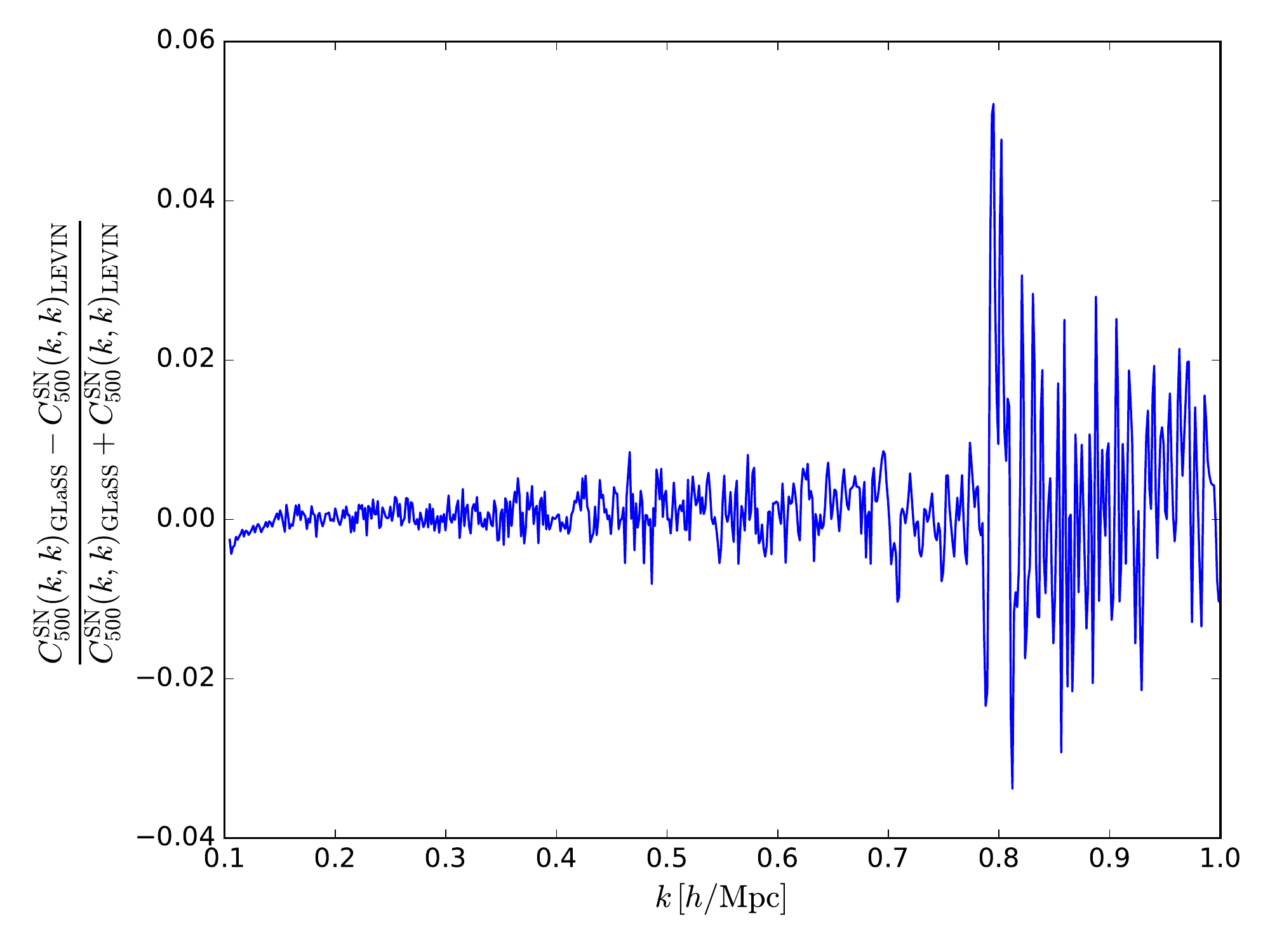}\par
\caption{Differences between the predictions for the signal (\textit{left} panels) and noise (\textit{right} panels) contributions to the covariance matrices for multipoles $\ell=100$ (\textit{top} panels) and $\ell=500$ (\textit{bottom} panels), normalised to their sum. We stress here again that the discrepancies at high $k$ should not be a concern because the $k$-regimes where they originate produce contributions very much subdominant with respect to the peaks of the signal curves, and also with respect to the relevant contributions from the noise (cf. Figs.~\ref{fig:comparison_signal}, \ref{fig:comparison_noise}). In the signal plots we shade the regions where the signal is a fraction $\leq 1/1000$ of the noise (cf. Fig.~\ref{fig:comparison_signal}): these regions correspond tho the $k$ values where the differences between the two codes are bigger, however since the signal contributions from these regions are negligeable, this discrepancy can be safely ignored.}
\label{fig:relativediff}
\end{figure*}

As a second diagnostic for our comparison, we consider individually the signal and noise contributions to the covariance matrices (Eqs.~\ref{eq:3Dcovariance} and \ref{eq:noise}, respectively) for two different multipoles, $\ell = 100$ and $\ell = 500$. For both signal and noise we compare the elements on the diagonal $C_{\ell}(k,k)$, and plot them respectively in Fig.~\ref{fig:comparison_signal} and Fig.~\ref{fig:comparison_noise}. In the noise case we also multiply the curves by $k^2$, to check that they effectively become flat as expected. The predictions show good agreement, with differences of at most a few percent (in the lower $k$ range for the signal, and over the entire $k$ range for the noise), as visible also from Fig.~\ref{fig:relativediff}, where we show the differences between the codes, normalised to the sum of their predictions. The disagreement in the signal plot towards the higher end of the $k$ range is due to the numerical noise present in the GLaSS computations; however, this discrepancy can be disregarded because the contributions from those $k$-regimes ($k \gtrsim 0.2 h/\mathrm{Mpc}$ for $\ell =100$, $k \gtrsim 0.4 h/\mathrm{Mpc}$ for $\ell =500$) are many orders of magnitude smaller than the main contributions around the peak of the curves, and also much smaller than contributions from the noise (cf. Fig~\ref{fig:comparison_noise}). For $\ell = 100$, the Levin and {\tt GLaSS} predictions coincide until approximately $k \simeq 0.2 h/\mathrm{Mpc}$: at this point the behaviour of the curve for {\tt GLaSS} starts being dominated by numerical noise, while the Levin signal decreases in a smoother way. The same happens for $\ell = 500$, but the disagreement starts at approximately $k \simeq 0.4 \, h/\mathrm{Mpc}$. In both cases however, the signal predictions in those $k$-regimes are at least 3-4 orders of magnitude smaller than the contributions around the peak of the curves, just before and after approximately $k \simeq 0.1 h/\mathrm{Mpc}$, respectively. Importantly, the values of the signal curves for those $k$-regimes are even smaller than the contributions from the noise, which dominates in that regime by many orders of magnitude. This means that for practical purposes we can safely ignore the contributions from those $k$-regimes where the codes are apparently in disagreement in their signal predictions. In Figs.~\ref{fig:comparison_signal},\ref{fig:comparison_noise}, and the left panels of Fig.~\ref{fig:relativediff} we demonstrate this point by shading the regions where the signal contribution represents a fraction $\leq 1/1000$ of the noise contribution at the same $k$. These regions turn out to be the same where the signal predictions of the two codes disagree, thus demonstrating that this discrepancy can be safely disregarded. In the bottom panel of Fig.~\ref{fig:comparison_signal} we plot the same comparison between the signal predictions produced by both two codes, with a linear scale on the $y$-axis instead of the logarithmic one used in the top panel; this is another way to appreciate how subdominant the contributions coming from the higher end of the $k$-range are with respect to the signal coming from the lower $k$-range.

It is interesting to note that the disagreement is practically only evident in the signal predictions, while the noise part is much less affected. This may be due to the increased number of matrix multiplications that need to be performed in the calculation of the signal with respect to the noise (cf. Eqs.~\ref{eq:3Dcovariance} and \ref{eq:noise}). The fact that the number of integrations to carry out for the noise is higher means, in the {\tt GLaSS} implementation, that more matrix multiplications are required and these are sensitive to the resolution in $k$. Additionally, since in the noise part of the covariance matrix there are no multiplications by Bessel functions, this may suggest that the spikes at high-$k$ in the signal may also be due to the Bessel function resolution breaking down (as explained in Sec.~\ref{sec:Glass}, in {\tt GLaSS} the Bessel functions $j_\ell (x)$ are precomputed in a look up table in $\ell$ and $x$).
The code implementing the Levin method sources the matter power spectrum from the Einstein-Boltzmann solver Cosmic Linear Anisotropy Solving System \citep[CLASS,][]{Blas2011}, while for this code comparison the matter power spectrum used by {\tt GLaSS} has been sourced from the Code for Anisotropies in the Microwave Background \citep[CAMB,][]{Lewis1999}. The CLASS and CAMB codes have been compared in their predictions \citep{Lesgourgues2011b}. Therefore, in comparing the Levin and {\tt GLaSS} methods, the matter power spectrum has been ruled out as a possible source of discrepancy. 

We conclude this section with a note on the performance of the two codes. The code implementing the Levin integration has been developed explicitly for the production of precise 3D cosmic shear cosmological forecasts and has been recently used to this purpose in \citet{SpurioMancini2018}. As one can see from Figs.~\ref{fig:comparison_signal} and \ref{fig:comparison_noise}, the curves produced with the Levin method are very smooth, showing the high precision achieved by the method. This compensates for the relatively low speed of the code, necessary to achieve that precision. {\tt GLaSS} on the other hand, has not been developed for 3D cosmic shear only; in \citet{Taylor2018a, Taylor2018b} it is introduced as a means to compute lensing spectra for arbitrary weighting functions and, importantly, for integration within the cosmological module {\tt Cosmosis}. This means that speed has been a crucial goal in developing the code and the method used for the matrix multiplications indeed allows for greater speed than the one achieved with the Levin method. However, numerical noise remains higher: to overcome this issue, one would need to increase the resolution at which the matrix multiplications are performed, but this would inevitably imply a slower performance of the code. We conclude that the use of the Levin or the {\tt GLaSS} method depends on the task to perform: if a high level of precision is required, the Levin method should be preferred, while if speed is a crucial requirement, {\tt GLaSS} can be a better option. For our purposes in this paper, i.e. the demonstration of a method for generating 3D lensing random fields on the sky and the calculation of Minkowski Functionals associated to these fields, both methods are equally valid for the computation of the 3D cosmic shear covariance matrices, which represent the starting point of the algorithms described in the following sections.

\section{Generation of spin-2 random fields on the sky}\label{sec:random_fields}
In this section we show how to generate random fields on the sky starting from the full 3D cosmic shear covariance matrix. As shown in Sec.~\ref{sec:3Dcosmicshear}, the full covariance matrix can be decomposed in $C_{\ell}(k, k')$ for each multipole $\ell$, given that the assumed isotropy of the shear field implies multipole independence $\left\langle \gamma_{\ell m}(k) \gamma_{\ell' m'}(k') \right\rangle = C_{\ell}(k, k') \delta_{\ell \ell'} \delta_{m m'}$. We detail our procedure considering for \mbox{simplicity} the convergence $\kappa$, as it is a scalar field and therefore easier to analyse. The convergence shares essentially the same covariance matrix with the shear field, each $\ell$-block only being rescaled by a prefactor $\frac{\ell (\ell + 1)}{(\ell+2)(\ell-1)}$ \citep{Castro2005} that plays a role only for the very largest angular scales. The generalisation to the spin-2 case for the shear field simply requires starting from the original shear covariance matrix and replacing the transforms from Fourier coefficients to configuration space with their spin-2 extensions. To demonstrate our field generation procedure we use the covariance matrices produced with the Levin integration.

Our aim is to generate modes of the convergence field in Fourier space $\kappa_{\ell m}$ and to transform them back into configuration space using the HEALPix \citep{Gorski2005} in-built function \texttt{alm2map}, in its scalar version for the convergence case (for the shear, one simply needs to activate the option \texttt{pol = True} that allows the user to deal with spin-2 fields). This way we can obtain samples of the convergence field in configuration space $\kappa(r, \theta, \phi)$, on spherical shells corresponding to different values of the radius; on each shell, the field can be discretised on a HEALPix map (an example of the final result is given in Fig.~\ref{fig:nested}). The procedure described in the following is similar to the one used in the code FLASK \citep{Xavier2016} to generate samples of the density, convergence and shear fields on redshift slices, starting from tomographic weak lensing covariance matrices $C_{i j}(\ell)$, where the indices $i$ and $j$ run over the redshift slices and the type of field (density, convergence or shear). In FLASK, the problem of generating correlated random fields across different redshift slices is dealt with by means of a Cholesky decomposition of the correlated covariance matrices ${C}_{i j}(\ell)$. The Cholesky decomposition rewrites the covariance as the product of an upper and lower triangular matrix. Here, the situation is similar in that we also have correlated multipoles belonging to the different radial slices, however the correlation is in terms of the wavector $k$ rather than the tomographic/field index $i$. This difference originates from the fact that we start from the 3D cosmic shear covariance matrices $C_\ell (k, k')$, as opposed to the tomographic $C_{i j} (\ell)$ matrices in FLASK. Additionally, in FLASK correlations between density, convergence and shear fields can be considered if the user desires, while here we concentrate singularly on the generation of convergence or shear fields and do not consider their cross-correlations. The fact that the random fields at different wavectors are correlated is ultimately due to the fact that the lensing field is not homogenenous along the line of sight, due to the mode-coupling effect of the lensing kernel, the source redshift distributions and the redshift error probability (cf.~Eqs.~\ref{eq:3Dcovariance}-\ref{eq:U}).

The assumption of statistical isotropy implies that modes $\kappa_{\ell m}$ of the convergence field at different multipoles $\ell$ and $m$ can be generated independently. The number of $\ell$ multipoles is in principle infinite, however practically there will be a $\ell_{\mathrm{max}}$ which sets the maximum resolution. We use $\ell_{\mathrm{max}} = 3 N_{\mathrm{side}}$, where $N_{\mathrm{side}}$ is a HEALPix parameter describing the resolution of the HEALPix grid \citep{Gorski2005}. The choice $\ell_{\mathrm{max}} = 3 N_{\mathrm{side}}$ is the same made by \citet{Lim2012} in their CMB analysis and guarantees that the grid size is comparable to the smallest angular scale considered, corresponding to $\ell_{\mathrm{max}}$. For each $\ell$ value, $m$ ranges from $-\ell$ to $+\ell$, so that there are $2\ell+1$ $m$ values for each multipole $\ell$. However, due to the hermiticity of the convergence field, we actually consider only $\ell+1$ modes from 0 to $\ell$.  
We employ a Cholesky decomposition of the covariance matrices to deal with the fact that modes corresponding to different $k$ values are correlated:
\begin{align}
C_{\ell} (k, k') = \sum_{p} T_{\ell} (k, p) T_{\ell} (p, k'),  
\end{align}
where $\mathbf{T}(\ell)$ are (lower) triangular matrices, which we can later use to generate correlated random variables $\kappa_{\ell m}(k)$, e.g. Gaussian distributed, 
\begin{align}
\kappa_{\ell m} (k) = \sum_{p} T_{\ell} (k, p)  \, n_{\ell m}(k),
\end{align}
where $n_{\ell m}(k)$ are independent, Gaussian distributed random variables with zero mean and unit variance. To obtain our convergence field samples in configuration space, we transform back from Fourier space, first by multiplying by a spherical Bessel function and $k^2$ and integrating over $k$, as indicated by Eq.~\ref{eq:sphericalFourier-Bessel}, and then acting with the HEALPix routine \texttt{alm2map} to obtain the field samples on a discretised grid in the angular coordinates. This way we are essentially performing in two steps an inverse spherical-Bessel transform.

We summarise schematically our procedure in Algorithm \ref{alg:RandomFields}. We implemented it in a Python routine, leveraging parallelisation on multiple cores with {\tt joblib}. The problem is embarassingly parallel, since the correlation of the fields on different radii is preserved by the starting cosmic shear covariance matrix, while different realisations of the random fields are completely independent. The fact that the covariance does not depend on $m$, but only on the multipole $\ell$, can be used to speed up calculations, as one needs to perform the Cholesky decomposition only once per each multipole $\ell$, and can then use the decomposition for all $m$'s pertaining to that $\ell$ mode.

\begin{algorithm}
\SetKwInOut{Input}{input}\SetKwInOut{Output}{output}\SetKwInOut{Method}{method}
  \Input{Covariance matrix (e.g. for the convergence) $\left\langle \kappa_{\ell m} \kappa_{\ell' m'} \right\rangle$ = $C_\ell (k, k') \delta_{\ell \ell'} \delta_{m m'}$}
  \Output{$
\kappa(r, \theta, \phi)$. For each fixed radius $r$ in $r_1, \dots r_{N_\chi}$, create a HEALPix map on discretised $\theta$ and $\phi$}
  \BlankLine
  \Method{
  $\forall r \in [r_1, \dots r_{N_\chi}]$: \\
	  \quad $\forall \ell \in [0, \ell_{\mathrm{max}}]$: \\
	  		\quad \quad       Cholesky decompose $\boldsymbol{C}_\ell=\boldsymbol{T}_\ell \boldsymbol{T}_\ell^T$\;
	   		\quad \quad $\forall m \in [0, \ell]$: \\
       		\quad \quad       sample $\boldsymbol{z}~\sim~\Norm({0},I)$\;
 	   		\quad \quad       $\kappa_{\ell m}(k)=\boldsymbol{T}_\ell \boldsymbol{z}$\;
 	   		\quad \quad 	  $\kappa_{\ell m}(k) \rightarrow \int \mathrm{d} k \, k^2 \, j_{\ell}(k r) \rightarrow \kappa_{\ell m} (r)$\;
 	   		\quad \quad 	   $\kappa_{\ell m}(r) \rightarrow$ HEALPix \texttt{alm2map} $\rightarrow \kappa(r, \theta, \phi) $
  	   		\BlankLine}
 \caption{Algorithm for generation of lensing Gaussian random fields on spherical shells}\label{alg:RandomFields}
\end{algorithm}

We show an example of 3D reconstruction on 3 slices in redshift (or equivalently, comoving distance) in Fig.~\ref{fig:nested}.

\section{Minkowski Functionals of scalar fields on the sphere}\label{sec:minkowski}
\begin{figure}
\centering
\includegraphics[scale=0.13, trim={21cm 0 0 0},clip]{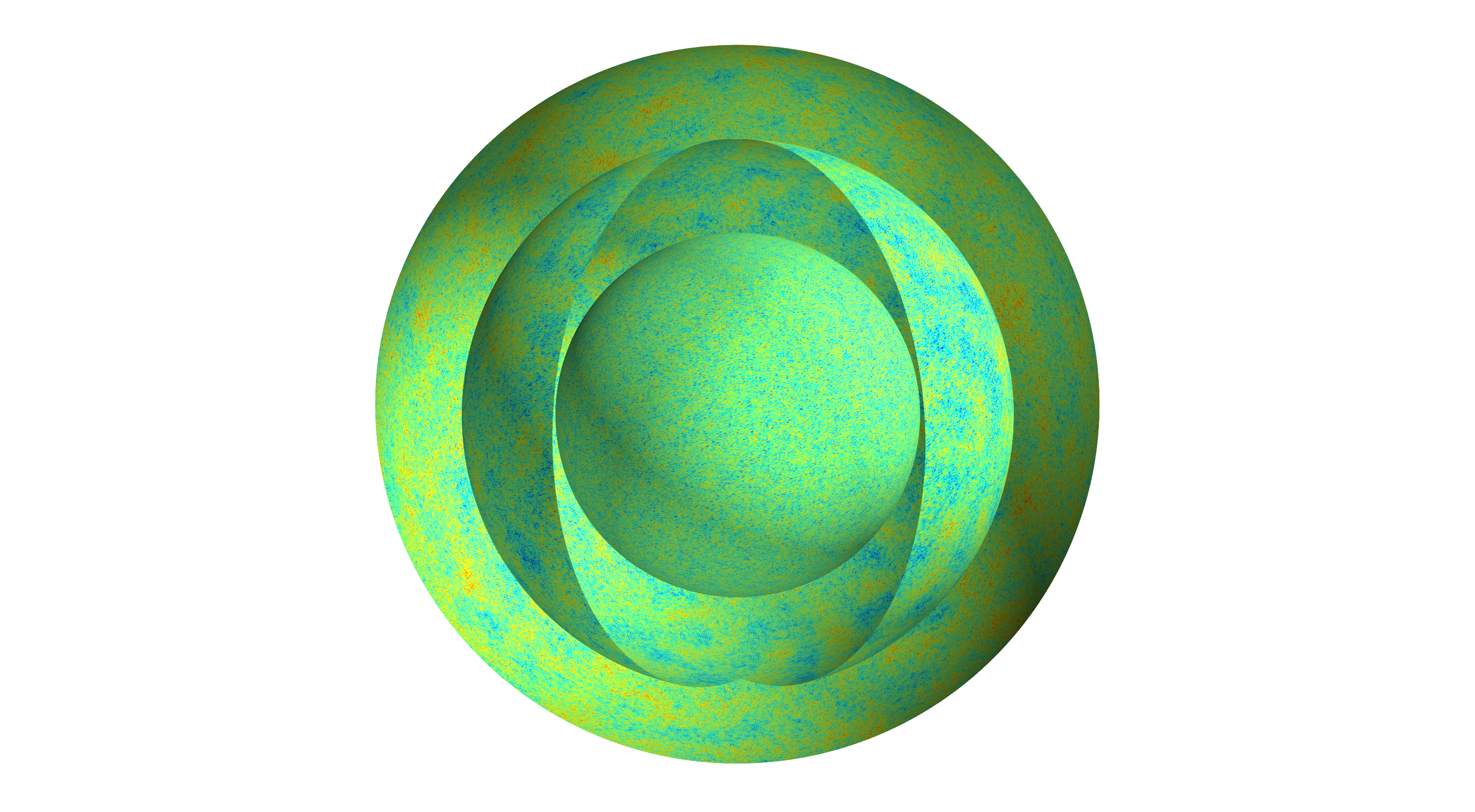}
\caption{Convergence field sampled at three different values of the radius $\chi$, with the observer situated in the centre. A section of the outer and middle sphere has been removed to facilitate visualisation. The lensing covariance matrix which we used for sampling the random field is given by Eq.\ref{eq:3Dcovariance}. We consider only contributions from the signal part of the covariance matrix, and use 30 $\ell$ modes ranging between 10 and 1000. We use a linear matter power spectrum for the calculation of the covariance.}
\label{fig:nested}
\end{figure}

In this section we briefly introduce Minkowski Functionals (MFs) and apply them to the generated random fields introduced in Sec.~\ref{sec:random_fields}. For Gaussian random fields MFs can be calculated analytically.  
We will compare this theoretical prediction with the MFs calculated directly from the HEALPix maps as a proof of concept. In particular we will calculate the MFs on spheres of different radii (compare Fig.~\ref{fig:nested}) and estimate the covariance between the different MFs at those radii. Repeating this whole procedure for different starting lensing covariance matrices (e.g varying each time one parameter), we can then produce a likelihood function dependent on the underlying cosmology. 

\subsection{General definition}
Here we define the MFs, concentrating on the aspects that are more interesting for cosmological applications and referring the reader to e.g. \citet{Mecke1994} for further mathematical details. In our definitions we follow the notation of \citet{Schmalzing1998} and \citet{Lim2012}. 
 
MFs are quantities that characterize the morphology of sets in an $n$ dimensional space $\mathcal{M}$. To be considered morphological, a quantity needs to be invariant under translation and rotations; while one could think that there are many such quantities, Hadwiger's theorem \citep{Hadwiger1957} states that in an $n$ dimensional space there exist only $n+1$ linearly independent morphological functionals, from which all the others can be derived. These are the so--called  Minkowski Functionals. On the 2-dimensional sphere, $\mathbb{S}_2$, there are $2 + 1 = 3$ MFs, carrying clear geometrical interpretations and in particular representing the area, circumference and integrated geodesic curvature of an excursion set, i.e. a region where the field exceeds some threshold level. Given a threshold $\nu$ and a smooth scalar field $u$, the excursion set $Q_\nu$ is mathematically defined as
\begin{align}
Q_\nu = \big\{ x \in \mathcal{M} \, | \, u(x) > \nu \big\}\; ,
\end{align}
while its boundary $\partial Q_\nu$ is given by
\begin{align}
\partial Q_\nu = \big\{ x \in \mathcal{M} \, | \, u(x) = \nu \big\}\; .
\end{align}
When considering the 2-sphere $\mathbb{S}_2$,
the first MF $V_0(\nu)$ can be interpreted as the area of $Q_\nu$, 
\begin{align}\label{eq:v0}
V_0(\nu) := \frac{1}{4\pi} \int_{\mathbb{S}_2} \mathrm{d} \Omega \, \Theta (u-\nu)\; ,
\end{align}
where $\Theta$ is the Heaviside function. The total length of the boundary of $Q_\nu$ gives the second MF
\begin{align}\label{eq:v1}
V_1(\nu) := \frac{1}{16 \pi} \int_{\partial \mathbb{S}_2} \mathrm{d} l = \frac{1}{16 \pi} \int_{\mathbb{S}_2} \mathrm{d} \Omega \, \delta(u-\nu) \, |\nabla u|\; .
\end{align}
Here $\delta$ is the delta distribution and $|\nabla u|$ is the norm of the gradient of $u$. Finally, the third MF is the integral of the
quantity $\kappa$ along the boundary
\begin{align}\label{eq:v2}
V_2(\nu):= \frac{1}{8 \pi^2} \int_{\partial \mathbb{S}_2} \mathrm{d} l \, \kappa = \frac{1}{8 \pi^2} \int_{\mathbb{S}_2} \mathrm{d} \Omega \, \delta(u-\nu) \, |\nabla u| \, \kappa\; ,
\end{align}
where $\kappa$ is the geodesic curvature: this describes how much a curve $\gamma$ is different from a straight line, i.e. from a geodetic. For a normalised tangent, i.e. $|\dot{\gamma}| = 1$, the curvature is defined through
\begin{align}\label{eq:k}
\kappa := |\nabla_{\dot{\gamma}} \dot{\gamma}|\; , 
\end{align}
where $\nabla_{\dot{\gamma}}$ represents the covariant derivative along the tangent vector $\dot{\gamma}$ of the curve.
\citet{Schmalzing1998} show how to calculate $\kappa$ on a generic manifold, which in the case of $\mathbb{S}_2$ reads  
\begin{align}
\kappa = \frac{2 u_{;\theta} u_{;\phi} u_{;\theta \phi} - u_{;\theta}^2u_{;\phi\phi} - u_{;\phi}^2u_{\theta\theta}}{u_{;\theta}^2+u_{;\phi}^2},
\end{align}
with the semicolon indicating a covariant derivative.

\begin{figure}
\centering
\includegraphics[width=0.5\textwidth]{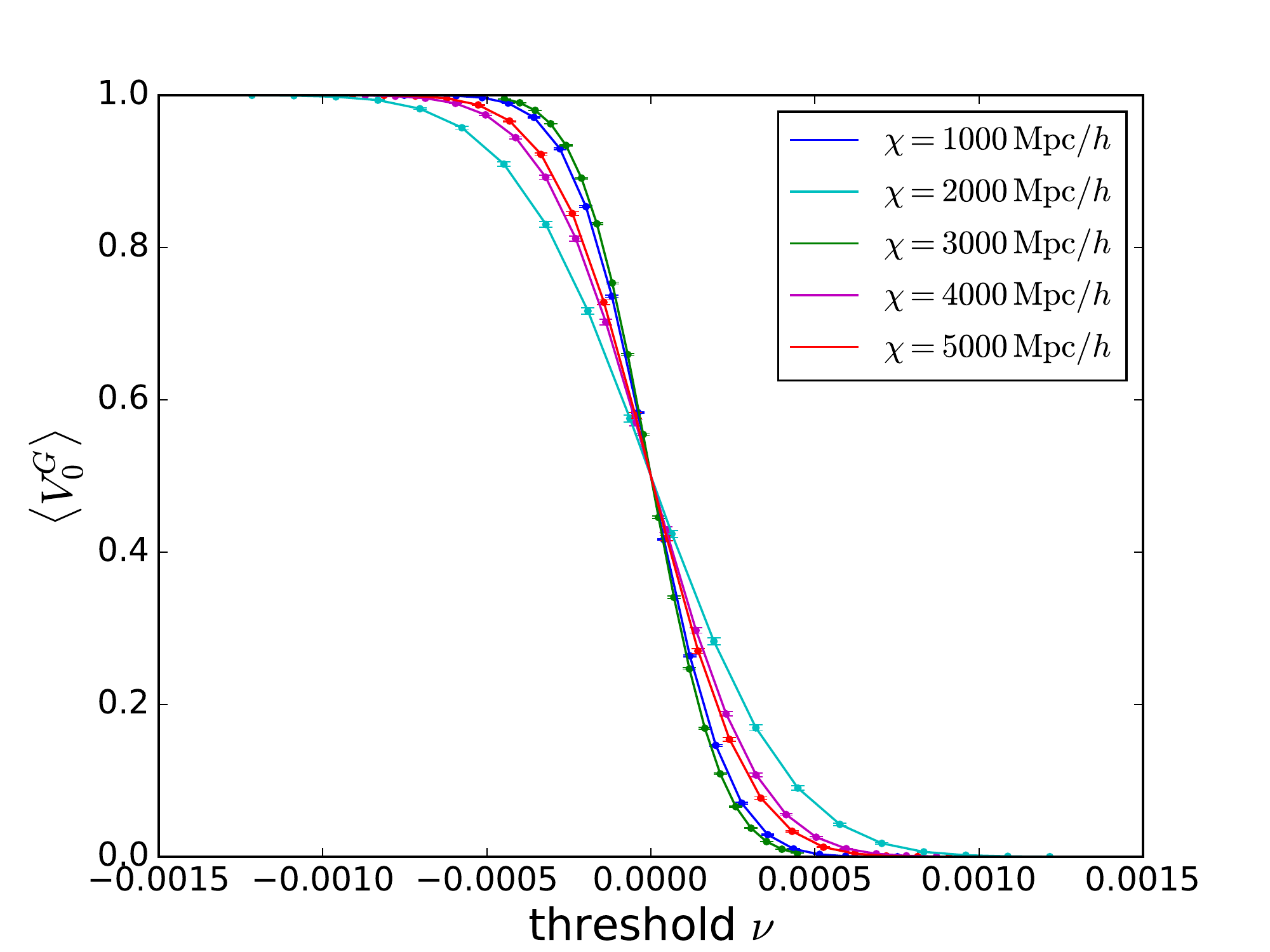}
\caption{Numerical estimations of the first MF $V_0^G$ (\textit{dots}), calculated on our generated Gaussian fields at different values of the radius (represented by different colours), compared with the theoretical predictions given by Eq.~\ref{eq:areatheory} (joined by \textit{lines}), as a function of the threshold $\nu$. The range of the thresholds always varies between $-4 \sqrt{\sigma}$ and $+4 \sqrt{\sigma}$, where $\sigma$ is the (average) variance of the lensing field at a fixed radius.}
\label{fig:area}
\end{figure}

\subsection{Numerical calculation of MFs}

Numerical estimates of the MFs calculated on our realisations of the lensing fields can be obtained using the software HEALPix \citep{Gorski2005}, as we explain in the following. We first generate full sky maps of e.g. the convergence field, on concentric spherical shells at different radii, starting from the 3D covariance matrices; to this purpose we follow the procedure described in sec.~\ref{sec:random_fields}. We then calculate numerically the MFs by directly implementing the integrals in Eqs.~\ref{eq:v0}-\ref{eq:v2}; our algorithm closely follows the one used in \citet{Schmalzing1998} and \citet{Lim2012} and is reported in the following.

\begin{figure}
\centering
\includegraphics[width=0.5\textwidth]{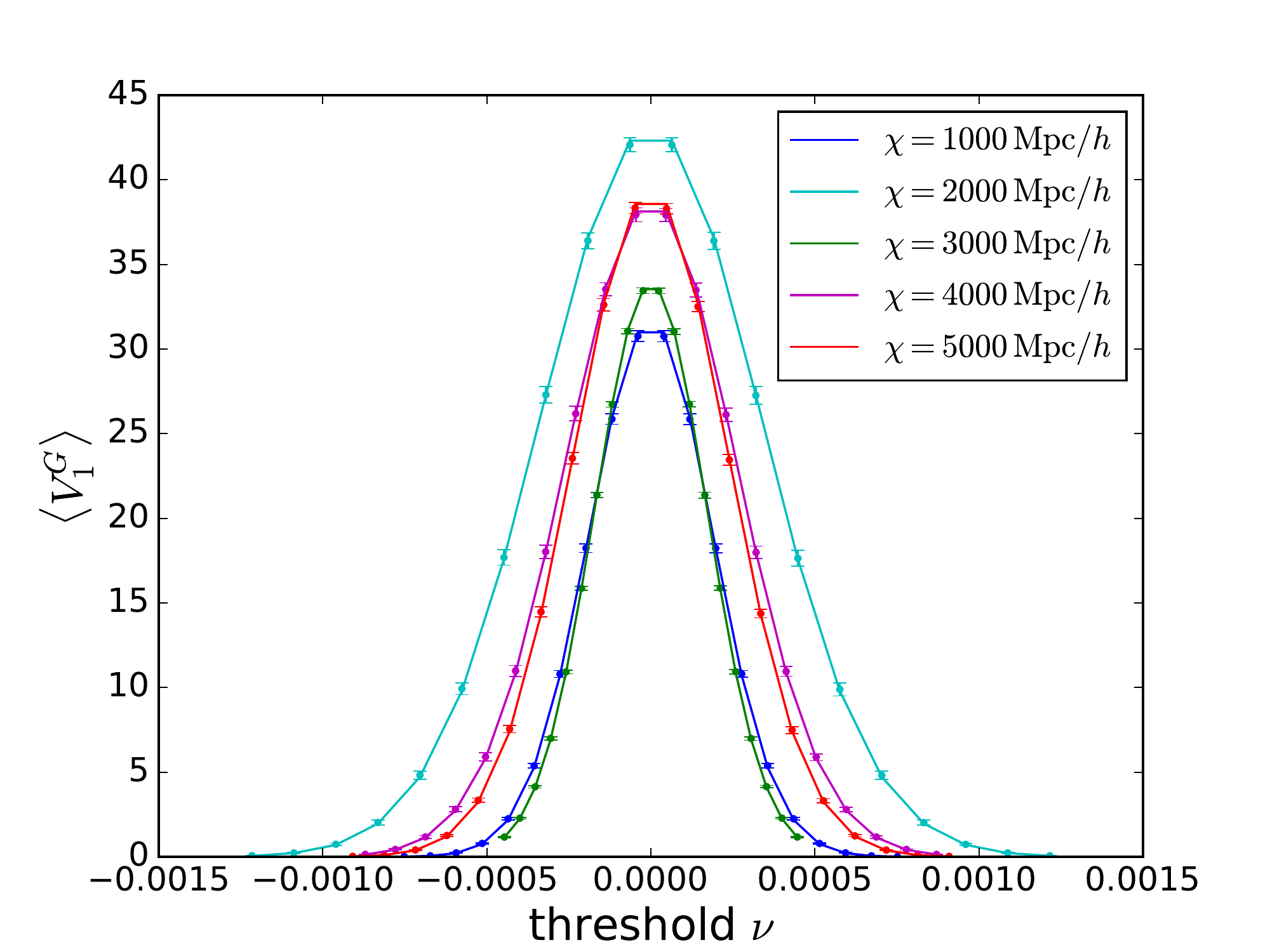}
\caption{Numerical estimations of the second MF $V_1^G$ (\textit{dots}), calculated on our generated Gaussian fields at different values of the radius (represented by different colours), compared with the theoretical predictions. The colour scheme is the same as in Fig.~\ref{fig:area}.}
\label{fig:perimeter}
\end{figure}

Given the values $u(x_i)$ of a field on a pixelated map, HEALPix provides useful routines that allow for the calculation of first and second partial derivatives at each pixel in $(\ell, m)$ spherical harmonic space. We use this to obtain the three numerical MFs for $\mathbb{S}_2$, which we label $V_i (i = 0,1,2)$, via a sum over all pixels
\begin{align}
V_i(\nu) := \frac{1}{N_{\mathrm{pix}}} \sum_{j = 1}^{N_{\mathrm{pix}}} \mathcal{I}_i (\nu, x_j)\; 
\end{align}
of the respective integrands
\begin{align}
\mathcal{I}_0 (\nu, x_j) &:= \Theta(u-\nu)\;, \\
\mathcal{I}_1 (\nu, x_j) &:= \frac{1}{4} \delta(u-\nu) \sqrt{u_{;\theta}^2+u_{;\phi}^2}\;, \\
\mathcal{I}_2 (\nu, x_j) &:= \frac{1}{2\pi} \delta(u-\nu) \frac{2 u_{;\theta} u_{;\phi} u_{;\theta \phi} - u_{;\theta}^2u_{;\phi\phi} - u_{;\phi}^2u_{\theta\theta}}{u_{;\theta}^2+u_{;\phi}^2}\;,
\end{align}
where the semicolon indicates again a covariant derivative, 
\begin{align}
u_{;\theta}&:=\partial_{\theta} u, \quad u_{;\phi}:=\frac{1}{\sin \theta} \partial_\phi u, \quad u_{;\phi\phi}:=\frac{1}{\sin^2 \theta} \partial_\phi^2 u + \frac{\cos \theta}{\sin \theta}  \partial_\theta u\;, \\
u_{;\theta \theta}&:=\partial_{\theta}^2 u, \quad u_{;\theta \phi}:=\frac{1}{\sin \theta} \partial_\theta \partial_\phi u - \frac{\cos \theta}{\sin^2 \theta} \partial_{\phi}u, \quad u_{;\theta}:=\frac{\partial}{\partial \theta}\;.
\end{align}
The integrands $I_1$ and $I_2$ involve the delta function: to approximate this numerically, \citet{Schmalzing1998} and \citet{Lim2012} use the Heaviside function
\begin{align}
\delta_N(x) := (\Delta \nu)^{-1} [\Theta(x+\Delta \nu/2) - \Theta(x-\Delta \nu/2)]\;.
\end{align}
This approximation of the delta function produces some numerical noise, which \citet{Lim2012} demonstrate to be due to the delta function discretization rather than some random noise which should disappear averaging over $n_R$ realisations. For our purposes, we do not consider the corrections proposed by \citet{Lim2012} to remove this discretisation effect and simply average over many realisations of the field. This is enough for our purposes, as our main goal is to test the field generation procedure rather than using the MFs to study e.g. non-Gaussianity as in \citet{Lim2012} (in which case these corrections should be taken into account). 

\begin{figure}
\centering
\includegraphics[width=0.5\textwidth]{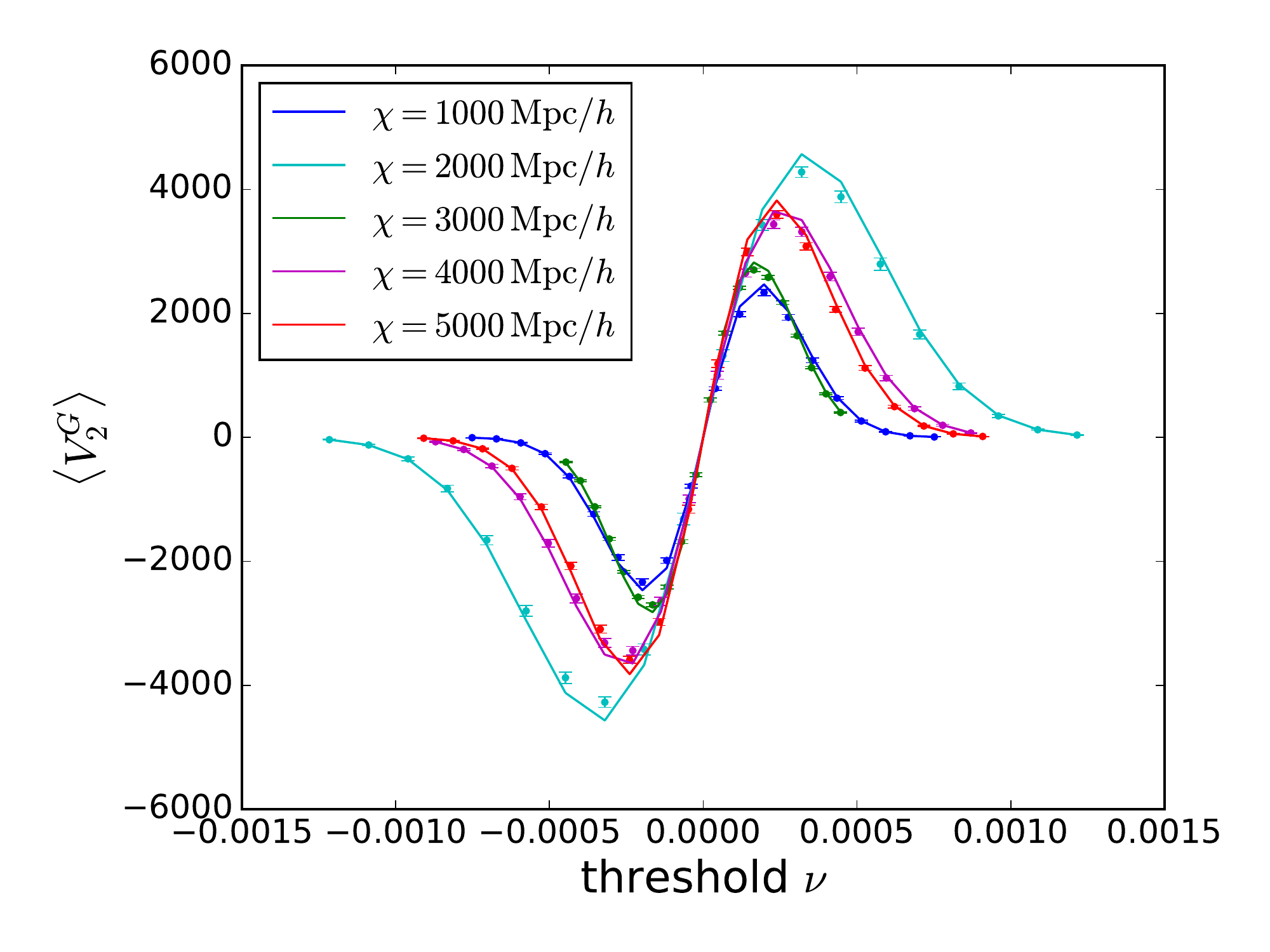}
\caption{Numerical estimations of the third MF $V_2^G$ (\textit{dots}), calculated on our generated Gaussian fields at different values of the radius (represented by different colours), compared with the theoretical predictions. The colour scheme is the same as in Fig.~\ref{fig:area}.}
\label{fig:euler}
\end{figure}

For Gaussian fields, as the ones we are considering here, the expectation values for the MFs are known analytically and equal to 
\begin{align}
\bar{V}_0^G (\nu) &:= \left\langle {V}_0^G (\nu) \right\rangle = \frac{1}{2} \left( 1 - \mathrm{erf} \left(\frac{\nu - \mu}{\sqrt{2\sigma}}\right) \right)\;, \label{eq:areatheory} \\ 
\bar{V}_1^G (\nu) &:= \left\langle {V}_1^G (\nu) \right\rangle = \frac{1}{8} \sqrt{\frac{\tau}{\sigma}} \mathrm{exp} \left( -\frac{(\nu - \mu)^2}{2 \sigma} \right)\;, \label{eq:perimetertheory} \\ 
\bar{V}_2^G (\nu) &:= \left\langle {V}_2^G (\nu) \right\rangle = \frac{1}{(2 \pi)^{3/2}} \frac{\tau}{\sigma} \frac{\nu - \mu}{\sqrt{\sigma}} \mathrm{exp} \left( -\frac{(\nu - \mu)^2}{2 \sigma} \right)\;. \label{eq:eulertheory}
\end{align}
Therefore we can compare our numerical estimates with the theoretical expectation values as a check for the validity of our field generation procedure. We perform this comparison in Figs.~\ref{fig:area},~\ref{fig:perimeter},~\ref{fig:euler}, where we overplot our numerical estimates and their expectation values. We consider all three MFs and show the comparison for five values of the radii, corresponding to five concentric shells over which we generate our lensing field. We calculate our MFs over a set of thresholds that always ranges between $-4 \sqrt{\sigma}$ and $+4 \sqrt{\sigma}$, where $\sigma$ is the variance of the lensing field at a certain radius. 
    
The error bars associated to our numerical estimates of the MFs are taken as the square root of the diagonal elements of the covariance matrix  of the MFs, computed as
\begin{align}\label{eq:covarianceMF}
\mathrm{Cov}_{ij} &= \frac{1}{n_R-1} \sum_{m=1}^{n_R} \Big( V_i^m - \left\langle V_i \right\rangle \Big) \left( V_j^m - \langle V_j \rangle \right), \nonumber \\ 
\quad &i,j = 0, \cdots 3 \cdot n_\chi \cdot n_\nu 
\end{align}
where the indices $i,j$ run over the type of Minkowski Functional (the three MFs $V_0, V_1, V_2$), the number of radii $n_\chi$ and the number of thresholds $n_\nu$. $\left\langle V_i \right\rangle$ denotes the mean of the MFs over all realisations $n_R$, $\left\langle V_i \right\rangle = \frac{1}{n_R} \sum_{m = 1}^{n_R} V_i^m$.
An example of this matrix is presented in Fig.~\ref{fig:covariance030}. We consider the covariance between all three MFs ($V_0$, $V_1$, $V_2$, each of them as a function of the threshold $\nu$, ranging from $\nu_1$ to $\nu_{\mathrm{max}}$), and we include the correlations between MFs belonging to each different radius (labelled by different $\chi$ value, from $\chi_1$ to $\chi_{\mathrm{max}}$). The values of the radii are the same used for Figs.~\ref{fig:area},~\ref{fig:perimeter},~\ref{fig:euler}. 
We stress here that the error bars depicted in Figs.~\ref{fig:area},~\ref{fig:perimeter},~\ref{fig:euler}, associated to the MFs calculated for each value of the threshold are not independent. Also, since the realisations of the random fields on different radial shells are not statistically independent, the MFs on different radii are not independent either.~To give a flavour of the correlations between the MFs, in Fig.~\ref{fig:correlation} we show the elements of the correlation matrix, i.e. the Pearson correlation coefficients $r_{ij}$ calculated as
\begin{align}\label{eq:pearson}
r_{ij} = \frac{\mathrm{Cov}_{i j}}{\sqrt{\mathrm{Cov}_{i i}}\sqrt{\mathrm{Cov}_{j j}}}, \quad &i,j = 0, \cdots 3 \cdot n_\chi \cdot n_\nu 
\end{align}
from the covariance matrix $\mathrm{Cov}_{ij}$ of our numerical estimates of the MFs, Eq.~\ref{eq:covarianceMF}. 
It follows from the Cauchy-Schwarz inequality that $-1 \leq r_{ij} \leq 1$. A value of $r_{i j} = 1$ indicates a perfect linear correlation between the two variables $i$ and $j$; a common interpretation is that in this case all data points in a sample lie on a straight line. This is also true if $r_{ij} = -1$, but the slope of the line is negative. A vanishing correlation coefficient implies that there is no linear correlation. If the correlation coefficient is positive, deviations of both variables from the mean tend to have the same sign, whereas opposite signs lead to a negative correlation coefficient.

As expected, we notice in particular a strong anti-correlation for $V_0$ centered around $\nu = 0$, as was expected by looking at Fig.~\ref{fig:area}. The same is true for $V_2$ (cf. Fig.~\ref{fig:euler}), while $V_1$ is strongly positively correlated (cf. Fig~\ref{fig:perimeter}). This high amount of (anti)correlation suggests that in the Gaussian case analysed here it is not necessary to consider a very high number of threshold values; however, this may not be true in the non-Gaussian case, where a higher resolution in the threshold values may be important to identify non-Gaussian features. Crucially important is, in all cases, a sufficient resolution in the HEALPix maps used at the beginning for the generation of the random fields, and later for the calculation of the MFs (in our estimates, we used the HEALPix parameters $N_{\mathrm{side}} = 256$ and $\ell_{\mathrm{max}} = 3 N_{\mathrm{side}} = 768$). This affects considerably the speed of the numerical implementation of these computations, however as mentioned earlier in Sec.~\ref{sec:random_fields} the generation of random fields and, separately, the calculation of the MFs (both happening at each realisation and at each radius) are embarrassingly parallel problems; this can be leveraged in practical implementations by employing parallelisation across multiple cores and nodes, without the need to worry about inter-process communication. 

\subsection{Inference from Minkowski functionals of Gaussian fields}

Introduced in cosmology by \citet{Mecke1994}, the main applications of MFs so far have been as probes of primordial non-Gaussianities \citep{Schmalzing1997, Winitzki1998, Schmalzing1998}, widely used in two and three dimensions, for instance on WMAP CMB data \citep{Hikage2008}, Planck CMB data \citep{Ducout2013, PlanckNG2014, PlanckNG2016, Novaes2016, Buchert2017} and on the SDSS galaxy catalogue \citep{Park2005, Hikage2006}. In the CMB case, MFs are suboptimal estimators of primordial non-Gaussianity parameters, while it has been shown that polyspectra provide minimum error bars for weak levels of non Gaussianity \citep{Babich2005}. Nevertheless, MFs constitute an attractive alternative to an analysis with polyspectra for a number of reasons: firstly, contrary to the bispectrum, MFs are defined in configuration rather than in Fourier space, so that a robust implementation for MFs becomes in practice easier to achieve; secondly, MFs are sensitive to the full hierarchy of higher order correlations, instead of third order only, and can provide additional information on all the non-linear coupling parameters $f_{NL}$, $g_{NL}$, ... which appear in the perturbative development of the primordial curvature perturbation \citep{Komatsu2001, Okamoto2002}; additionally, MFs can be analytically determined for Gaussian random fields; lastly, they are additive, which makes accounting for complicated survey geometries much easier compared to estimators of polyspectra.

In this work we propose (for the first time, as to our knowledge) MFs as an alternative probe of Gaussianity, in addition to non Gaussianity, in the sense specified in the following. We show how, assuming our MFs to be Gaussian distributed, we can use the MFs to probe the cosmology dependence of the fields realisations. This can be leveraged in future work to develop a full cosmological inference process based on the MFs calculated on lensing fields, of which we provide a first example here.

From a Bayesian perspective, assuming that our likelihood $L(V_i | \Omega)$ (the probability of having MFs $V_i$ given the cosmological parameters $\Omega$) is Gaussian is equivalent, considering a flat prior $p(\Omega)$ on the cosmological parameters $\Omega$, to having a Gaussian posterior $p(\Omega | V_i)$, since by virtue of Bayes theorem
\begin{align}
p(V_i | \Omega) \propto L(\Omega |  V_i) p(\Omega).
\end{align}
It follows that we are allowed to consider the likelihood and the posterior equivalently. 
In the Gaussian case, defining $\mathcal{L} = -\mathrm{ln} L$ and ignoring additive constants, we have that $-2 \mathcal{L} = \chi^2$, where the $\chi^2$ can be evaluated as
\begin{align}\label{eq:chisquare}
\chi^2 (\theta) = \sum_{i, j = 1}^{n_\nu \cdot n_\chi} \Bigg( \langle V_i(\theta) \rangle -  \langle V_i(\theta_0) \rangle \Bigg) \; \mathrm{cov}^{-1}_{ij} \left(\theta_0 \right) \Bigg( \langle V_j(\theta) \rangle -  \langle V_j(\theta_0) \rangle \Bigg),
\end{align}
where the averages are performed over the number of realisations $n_R$, while the indices $i,j$ run over the length of our data vector, i.e. we consider the MFs evaluated at all the $n_\nu$ thresholds and all the $n_\chi$ radii. The MFs depend on the cosmological parameters and so does the covariance matrix; for the calculation of the chi-square, we use the inverse evaluated at the fiducial model $\theta_0$.

We calculate the $\chi^2$ statistics with MFs obtained from the realisations of the lensing random fields (we consider the convergence in this example) at different values of one cosmological parameter, for simplicity. We consider 11 values of $\Omega_\mathrm{m}$, ranging from 0.25 to 0.35 in equidistant intervals of 0.01 centered on the fiducial value of 0.3. For each of the $\Omega_{\mathrm{m}}$ values we produce our 3D cosmic shear covariance matrix following the equations in Sec.~\ref{sec:3Dcosmicshear}, with either the Levin or the {\tt GLaSS} method. Once the full lensing covariance matrix is available, we use it to generate, according to the procedure described in Sec.~\ref{sec:random_fields}, $n_R$ realisations of the convergence field at $r_{N_\chi}$ values of the radius in configuration space. On each shell and for each realisation we also calculate the associated MFs, and store them in memory. Subsequently we use them to build the full covariance matrix, exactly as the one shown in the previous subsection, however this time we will have one covariance matrix of the MFs for each starting value of $\Omega_{\mathrm{m}}$. Inverting the covariance corresponding to our fiducial value $\Omega_{\mathrm{m}}=0.3$, we can then use it to calculate the $\chi^2$ following Eq.~\ref{eq:chisquare}.

The calculation of this inverse covariance matrix poses a numerical problem, in that its entries are very small and standard methods such as Gaussian elimination fail in producing a sensible inverse. We use therefore a Moore-Penrose pseudo-inverse matrix \citep{Dresden1920, Penrose1955}, after checking that it effectively produces an inverse covariance matrix that, multiplied by the covariance, gives back the identity matrix to within numerical precision. 

\begin{figure*}
\centering
\includegraphics[width=.95\textwidth,trim={0 5cm 0 0},clip]{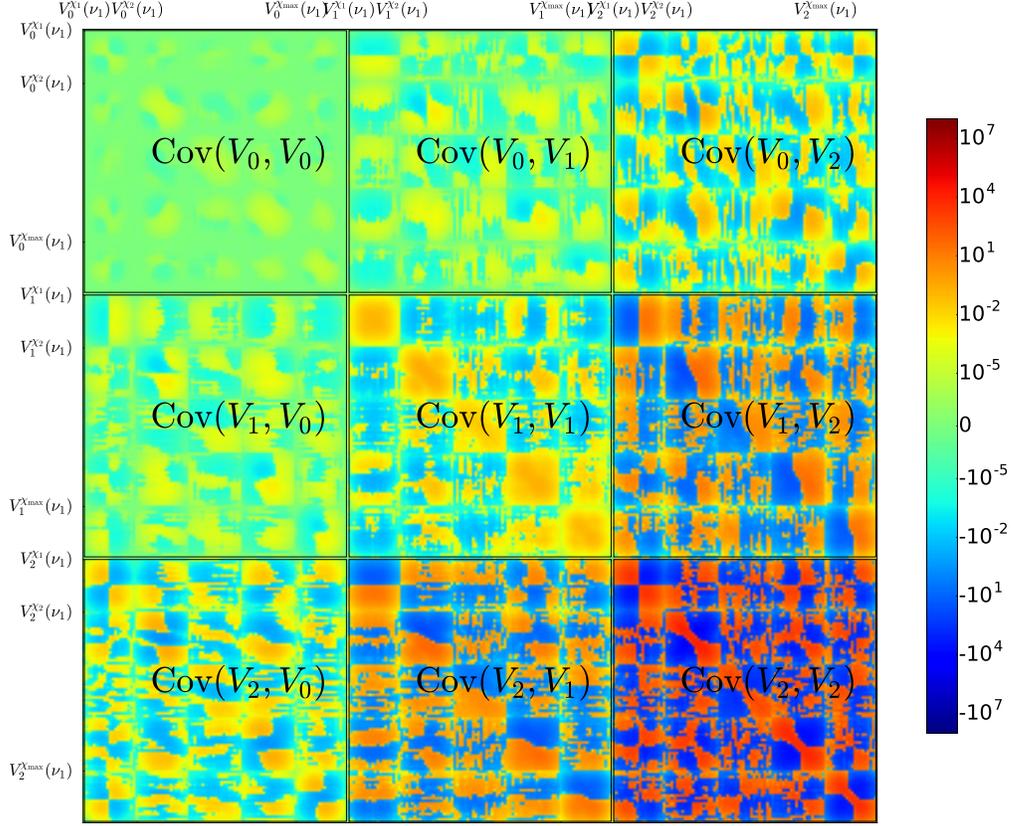}
\caption{Covariance matrix between different MFs at different radii. We consider correlations between all three MFs $V_0$, $V_1$, $V_2$, all functions of the threshold $\nu$ (ranging from $\nu_1$ to $\nu_{\mathrm{max}}$), as calculated at different radii (labelled by different $\chi$ values, ranging from $\chi_1$ to $\chi_{\mathrm{max}}$ and specifically equal to 1000, 2000, 3000, 4000 and 5000 Mpc/$h$, as in Figs.~\ref{fig:area}, \ref{fig:perimeter}, \ref{fig:euler}). In the matrix we indicate the block sub-matrices that represent the covariance between the three MFs. We used a logarithmic scale for both positive and negative values to highlight the many orders of magnitude spanned by the entries of the matrix and the different contibutions given by the three MFs.}
\label{fig:covariance030}
\end{figure*}

We calculate the $\chi^2$ isolating the different MFs in our data vector. This implies isolating from the full covariance matrix the relevant sub-blocks for the auto-correlation of $V_0, V_1$ and $V_2$ (which we will in the following schematically indicate with $\left\langle V_0, V_0 \right\rangle, \left\langle V_1, V_1 \right\rangle, \left\langle V_2, V_2 \right\rangle$, or $\mathrm{Cov}(V_0, V_0), \,  \mathrm{Cov} (V_1, V_1), \, \mathrm{Cov} (V_2, V_2)$). These sub-blocks can be visualised by looking at the corresponding sub-blocks in the covariance matrix plotted in Fig.\ref{fig:covariance030} (e.g. the correlation $\left \langle V_0, V_0 \right\rangle$ isolates the top left corner block); in Fig.~\ref{fig:chifull} we plot the $\chi^2$ curves obtained with the three MFs. We notice how the $\chi^2$ increases going from $V_0$ to $V_2$.

\section{Conclusions}\label{sec:conclusions}

\begin{figure*}
\centering
\includegraphics[width=.95\textwidth,trim={0 5cm 0 0},clip]{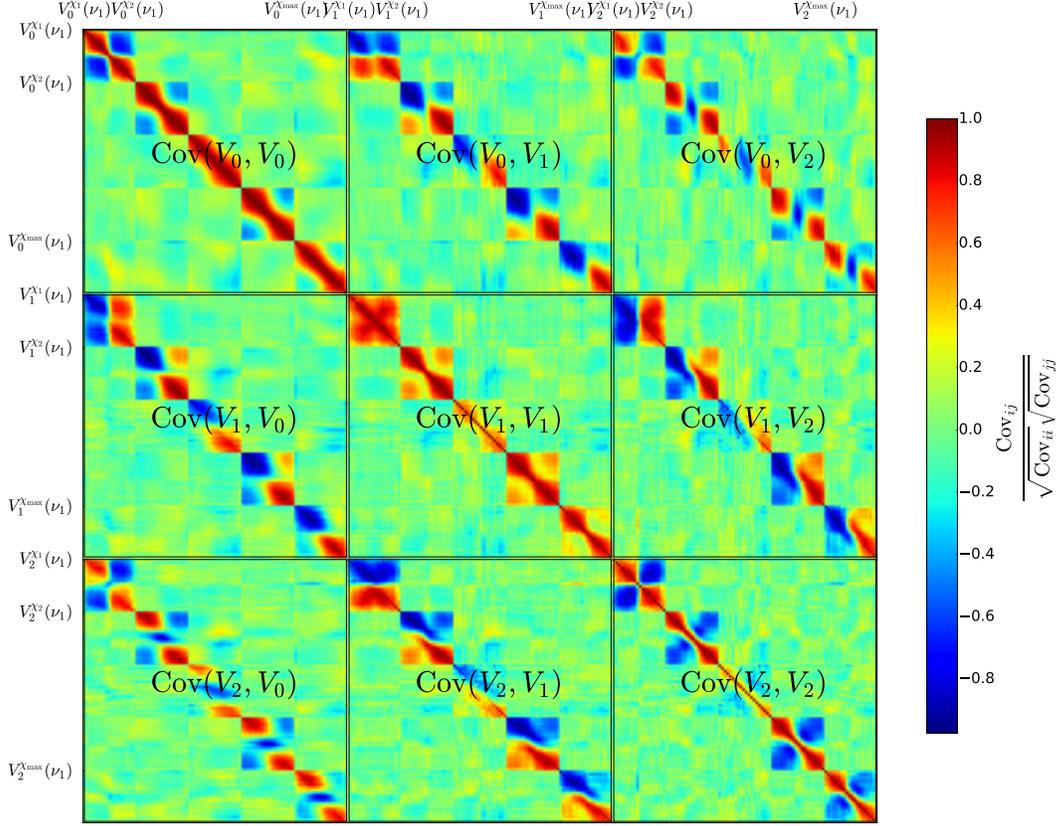}
\caption{Correlation matrix between different MFs at different radii; the matrix entries represent the Pearson correlation coefficient, obtained from the covariance matrix entries (the same plotted in in Fig.~\ref{fig:covariance030}) following Eq.~\ref{eq:pearson}. We consider correlations between all three MFs $V_0$, $V_1$, $V_2$, all functions of the threshold $\nu$ (ranging from $\nu_1$ to $\nu_{\mathrm{max}}$), as calculated at different radii (labelled by different $\chi$ values, ranging from $\chi_1$ to $\chi_{\mathrm{max}}$ and specifically equal to 1000, 2000, 3000, 4000 and 5000 Mpc/$h$, as in Figs.~\ref{fig:area}, \ref{fig:perimeter}, \ref{fig:euler}). In the matrix we indicate the block sub-matrices that represent the correlation between the three MFs.}
\label{fig:correlation}
\end{figure*}

\begin{figure*}
\centering
\includegraphics[width=0.7\textwidth]{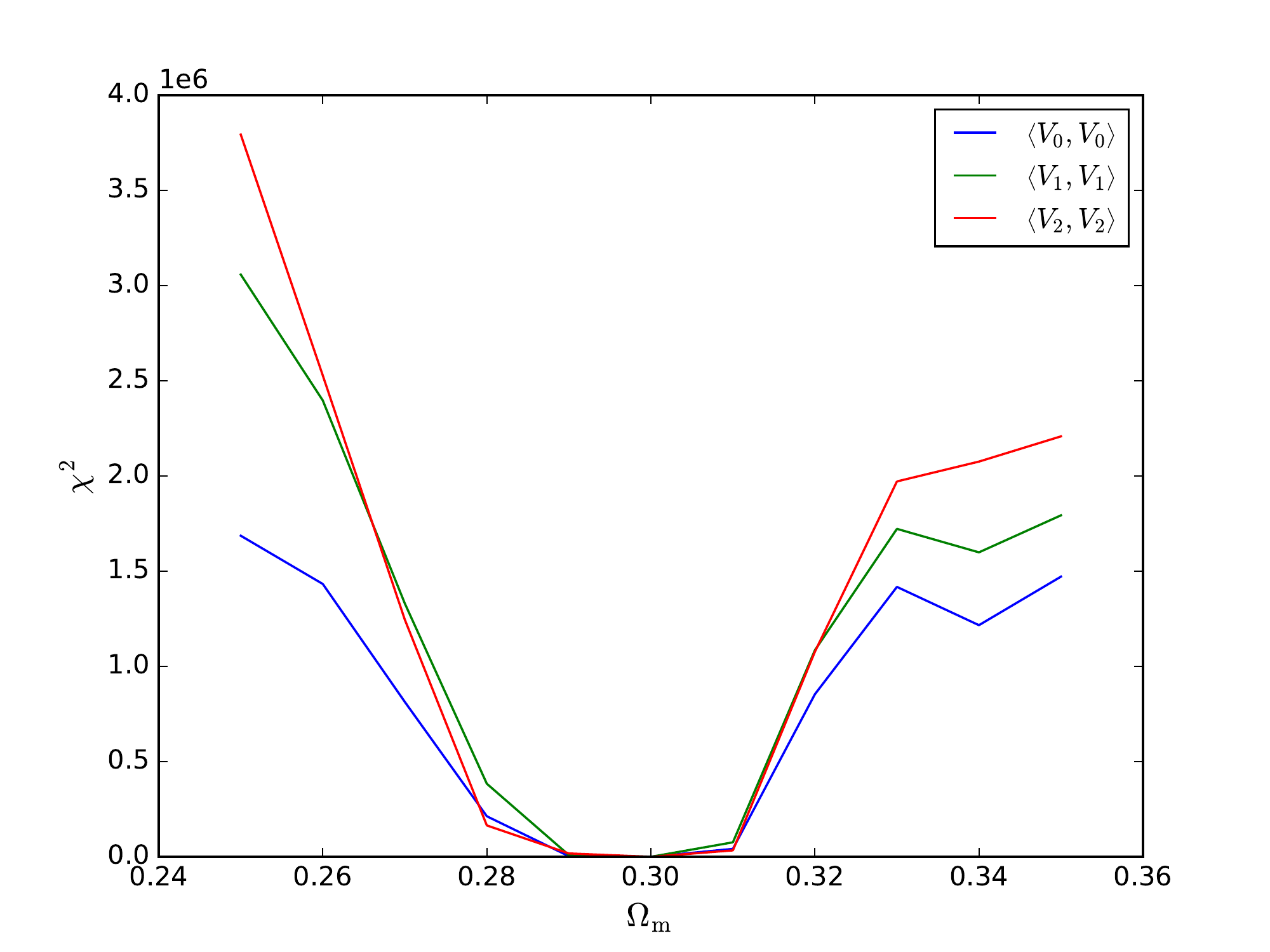}
\caption{$\chi^2$ obtained considering the covariance of different combinations of MFs, i.e. considering the three MFs singularly ($\left\langle V_0, V_0 \right\rangle$ (\textit{blue}), $\left\langle V_1, V_1 \right\rangle$ (\textit{green}) and $\left\langle V_2, V_2 \right\rangle$ (\textit{red})). Our fiducial model is represented by the choice $\Omega_{\mathrm{m}}=0.3$.}
\label{fig:chifull}
\end{figure*}


3D cosmic shear constitutes an alternative to a traditional tomographic analysis of a cosmic shear survey. The spherical-Bessel expansion of the shear field at the core of its formalism maximises the amount of redshift information; however, the calculation of the covariance matrices presents numerical difficulties due to the numerous integrations over highly oscillatory functions.

In this paper we described and compared two methods for the calculation of simulated 3D cosmic shear covariance matrices. While the first method implements the Levin technique for integration of the periodic oscillations of the Bessel functions, the second method, implemented in the code {\tt GLaSS}, tackles the integrations by matrix multiplications and appropriate use of the Limber approximation.  

We first compared the predictions of the two codes in terms of covariance matrices and found excellent agreement. We compared the output of the codes both in terms of the total signal-to-noise ratio and the single contributions to the covariance matrices $C_\ell$, for two different values of the multipole $\ell$, for both the signal and noise parts.

Once tested the accuracy of the predictions for the covariance matrices, we used the simulated matrices to generate Gaussian lensing fields on the sky. The procedure we described, based on a Cholesky decomposition of the $C_\ell$ matrices, allowed us to generate correlated Gaussian fields at different slices in comoving distance. We remark here that in our formalism for 3D cosmic shear we ignored complications arising from survey masks, assuming full sky coverage. Masked data can be readily accounted for by applying a mixing matrix/pseudo-$C_\ell$ like formalism. For 3D cosmic shear this is described in \citet{Kitching2014}, where it was applied to data. Both methods for the computation of 3D cosmic shear covariance matrices studied in this paper can be trivially modified to include such an effect. In our work we did not investigate masked data and the ability of a pseudo-$C_\ell$ method to produce power spectra that account for a mask; however, thanks to our method to generate 3D cosmic shear data, we could now test this pseudo-$C_\ell$ approach by masking the simulated data and we will investigate this in a future paper. 

The generation of normal and lognormal fields (the latter being easily obtainable from the former, by exponentiation of the Gaussian maps) can be used in future work to compute a realistic covariance matrix for a full 3D cosmic shear likelihood analysis. This should improve upon e.g. the CFHTLenS analysis for 3D cosmic shear \citep{Kitching2014}, where a covariance implementation similar to {\tt GLaSS} was used. \citet{Kitching2014} constructed a likelihood, in which the parameter dependency was in the covariance rather than the mean shear transform coefficients. This could be improved by having a likelihood in which the covariance is used as the mean and the 4-point covariance of the covariance used. 


The generation of normal and lognormal random fields, starting from the 3D cosmic shear covariance matrices, also constitutes the first step for the development of a Bayesian Hierarchical Model for 3D cosmic shear power spectra estimation \citep[following e.g. the work of][and extending it to a spherical-Bessel formalism]{Alsing2016, Alsing2017}, which can be investigated in future work. 

In this work we tested our random field generation procedure by calculating Minkowski Functionals associated to our Gaussian random fields and comparing them with their known expectation values. We found good agreement between our numerical estimates and their theoretical expectation values. We calculated our Minkowski Functionals separately on each spherical shell, however we stress here that the realisations of the random fields on different radial shells are not statistically independent, as one can appreciate from the correlation matrix presented in Fig.~\ref{fig:correlation}.~Future work should concentrate on estimating the full correlation between the Minkowski Functionals at different values of the radii, implementing a fully three-dimensional approach for their calculation \citep[see e.g.][for examples of Minkowski Functionals in 3D]{Hikage2003, Gleser2006, Yoshiura2017, Appleby2018}. Producing fully 3D Minkowski functionals for a lognormal field in 3D can be used in particular to extract non-Gaussian information from the shear field. 

Finally, we showed how Minkowski Functionals can also be used to extract Gaussian information by means of a likelihood analysis. We show an example of this in Fig.~\ref{fig:chifull}, where we plot the $\chi^2$ obtained from the covariance of the different Minkowski Functionals as a function of the varying cosmological parameter $\Omega_m$. This is a first example of a full cosmological inference process, making use of the Minkowski Functionals, that we plan to develop in future work. 

\section*{Acknowledgements}
\addcontentsline{toc}{section}{Acknowledgements}

ASM and RR acknowledge financial support from the graduate college \textit{Astrophysics of cosmological probes of gravity} by Landesgraduiertenakademie Baden-W\"urttemberg. PLT is supported by the UK Science and Technology Facilities Council. TK is supported by a Royal Society University Research Fellowship. The authors thank the German DFG Excellence Initiative for funding a mobility program, through the grant \textit{Gravity on the largest scales and the cosmic large-scale structure}, within which this work originated.

\bibliographystyle{plainnat}
\bibliography{references.bib} 


\appendix

\section{Shot noise in 3D cosmic shear}\label{app:noise}
Here we derive explicitly the expression for the shot noise contribution to the 3D cosmic shear covariance matrix, as given by Eq.~\ref{eq:noise}. The noise contribution is present in all the literature on 3D cosmic shear \citep[see e.g. the seminal papers][]{Heavens2003, Heavens2006}, however in the context of our code comparison for 3D cosmic shear it may arise difficulties due to the different conventions used for the spherical-Bessel formalism. We refer the reader also to Appendix C in \citet{Lanusse2015}, where a derivation of the shot noise term for 3D galaxy clustering is presented.

Shot noise arises by discretising the survey in cells that either contain one or zero galaxies \citep{Peebles1980}. We will keep the discussion more general here, for a random field $f(\vec{x})$ that is discretised on our series of cells labelled by index $i$. We will later specialise to our intrinsic ellipticity field. $n_i$ represents the occupation number of the cell and $f_i$ the value of the field in cell $i$:
\begin{align}
f(\vec{x}) = \sum_i \delta(\vec{x} - \vec{x}_i) \, n_i \, f_i.
\end{align}
We calculate the correlation (where $V$ is the ``volume'' factor for our field):
\begin{align}
\left\langle f(\vec{x}) f(\vec{x}) \right\rangle &= \sum_{i , j} \left\langle \delta^D(\vec{x} - \vec{x}_i) \delta^D(\vec{x} - \vec{x}_i) \, n_i \,  n_j \, f_i \, f_j \right\rangle \frac{1}{V^2} \\
&= \sum_{i, j} \delta^D(\vec{x}_i - \vec{x}_j) \left\langle n_i^2 \right\rangle \left\langle f_i^2 \right\rangle \frac{1}{V^2}\\
&= \sum_{i} \delta^D(0) \left\langle n_i \right\rangle \left\langle f_i^2 \right\rangle \frac{1}{V^2}\\
&= \sum_i\langle n_i \rangle \frac{\langle f_i^2\rangle}{V}
\end{align}
where we used the fact that $n_i$ and $f_i$ are uncorrelated, $\langle n_i^2\rangle = \langle n_i\rangle$ due to Poisson sampling and we assumed that only equal cells are correlated. In the last step we used that $\delta^D(0) = V = 4\pi$.
In our case, the random field we consider is the intrinsic ellipticity of the galaxies $\epsilon_S$. This is because, as already mentioned in Sec.~\ref{sec:3Dcosmicshear}, we assume the observed ellipticity $\epsilon$ to be the sum of the shear $\gamma$ and the intrinsic ellipticity $\epsilon_S$, and neglect correlations between $\gamma$ and $\epsilon_S$ as given by intrinsic alignments. We denote the intrinsic ellipticity dispersion as $\sigma_\epsilon$ (with a typical value $\sigma_\epsilon\simeq 0.3$). Expressing the field $f$ in a spherical basis as
\begin{align}
f_{\ell m} (k) = \sqrt{\frac{2}{\pi}}\sum_i \, n_i \, f_i \, j_\ell (k\chi_i) \, Y_{\ell m} \, (\mathbf{\hat{n}_i}),
\end{align}
and taking into account the redshift distribution of galaxies, one arrives at Eq.~\ref{eq:noise}.

\label{lastpage}

\end{document}